\documentclass[aps,
	prd, 
	onecolumn, 
	a4paper, 
	10pt, 
	preprintnumbers,  
	tightenlines, 
	notitlepage,
	floatfix,
	nofootinbib,
	longbibliographyprd,
	superscriptaddress]{revtex4-1}

\input{CQGformat.input}
\newcommand{\mr}{\ms\hline\ms}
\usepackage{etoolbox}
 
\makeatletter
\def\@mkboth#1#2{}
\newlength\appendixwidth
\preto\appendix{\addtocontents{toc}{\protect\patchl@section}}
\newcommand{\patchl@section}{%
  \settowidth{\appendixwidth}{\textbf{Appendix }}%
  \addtolength{\appendixwidth}{1.5em}%
  \patchcmd{\l@section}{1.5em}{\appendixwidth}{}{\ddt}%
}
\makeatother

\usepackage{IEEEtrantools}
\usepackage{microtype}
\usepackage[toc,title,page,titletoc]{appendix}
\usepackage{amsmath}
\usepackage{amssymb}
\usepackage{amsthm}
\usepackage{mathrsfs}
\usepackage{enumitem}
\usepackage[T1]{fontenc}
\usepackage{xfrac}
\usepackage{lmodern}
\usepackage[pdftex,breaklinks,colorlinks,
linkcolor=Blue,
citecolor=teal,
anchorcolor=red,
urlcolor=cyan]{hyperref}
\usepackage{mathbbol}
\usepackage[dvipsnames]{xcolor}
\usepackage[graphicx]{realboxes}
\usepackage{adjustbox}

\newcommand{\GHPw}[2]{\left\{ #1, #2 \right\}}

\renewcommand{\S}{{\mathcal S}}
\newcommand{\E}{{\mathcal E}}
\newcommand{\T}{{\mathcal T}}

\renewcommand{\O}{{\mathcal O}}

\newcommand{\sI}{{\mathscr I}}
\newcommand{\sH}{{\mathscr H}}

\usepackage[normalem]{ulem}
 
\newcommand{\emm}{m}
\newcommand{\ess}{s}

\font\ec=ecrm0800 at 12pt
\def\th{\hbox{\ec\char'336}}

\def\mb{\bar m}

\def\pheq{\phantom{=}}

\DeclareMathOperator{\thorn}{\text{\rm \th}}
\let\eth\relax
\DeclareMathOperator{\eth}{\text{\rm \dh}}

\newtheorem{theorem}{Theorem}

\newtheorem{lemma}[theorem]{Lemma}

\renewcommand{\Re}{\operatorname{Re}}
\renewcommand{\Im}{\operatorname{Im}}

\newcommand{\preindex}[2]{\,{}_{#2}#1}
\newcommand{\half}{\dfrac{1}{2}}

\newcommand{\intd}[1]{\textrm{d} #1}

\newcommand{\dd}{{\rm d}}

\def\ben{\begin{equation}}
\def\een{\end{equation}}
\def\bena{\begin{eqnarray}}
\def\eena{\end{eqnarray}}

\newcommand{\Y}[1]{\, {}_{#1}Y^{\phantom{\textrm{p}}}_{\ell, \emm}}

\newcommand{\Ylm}[3]{\, {}_{#1}Y_{#2, #3}}

\begin{document}

\title{Spin-2 Green's Functions on Kerr in Radiation Gauge}

\author{Marc Casals\footnote{Corresponding author.}$^{1,2,3}$, Stefan Hollands$^{1,4}$, Adam Pound$^5$, and Vahid Toomani}
%\email{marc.casals@uni-leipzig.de}

\address{$^1$ Institut f\"ur Theoretische Physik, Universit\"at Leipzig, Br\"uderstra{\ss}e 16, 04103 Leipzig, Germany.}
\address{$^2$ School of Mathematics and Statistics, University College Dublin, Belfield, Dublin 4, Ireland.}
\address{$^3$ Centro Brasileiro de Pesquisas F\'isicas (CBPF), Rio de Janeiro, 
CEP 22290-180, 
Brazil.}

%\email{stefan.hollands@uni-leipzig.de}
\address{$^4$ Max-Planck-Institute MiS, Inselstrasse 22, 04103 Leipzig, Germany.}

%\author{Adam Pound}
%\email{a.pound@soton.ac.uk}
\address{$^5$ School of Mathematical Sciences and STAG Research Centre, University of Southampton, Southampton SO17 1BJ, United Kingdom.}

%\author{Vahid Toomani}
%\email{vahidtoomani2002@gmail.com}

\ead{marc.casals@uni-leipzig.de, stefan.hollands@uni-leipzig.de, a.pound@soton.ac.uk, vahidtoomani2002@gmail.com}

\date{\today}

\begin{abstract}
We construct retarded and advanced Green's functions for gravitational  perturbations in Kerr in an ingoing radiation gauge. Our Green's functions have a frequency domain piece that has previously been obtained by Ori [Phys. Rev. D \textbf{67} (2003)] based on the Chrzanowski-Cohen-Kegeles metric reconstruction method. As is well known, this piece by itself is not sufficient to obtain an actual Green's function. We show how to complete it with a piece based on a method by Green et al. [Class. Quant. Grav. \textbf{37} (2020)]. The completion piece has a completely explicit form in the time-domain and is supported on pairs of points on the same outgoing principal null geodesic which are in the appropriate causal order. We expect our Green's functions to be useful for gravitational self-force calculations and other perturbation problems on Kerr spacetime.
\end{abstract}

\maketitle
\tableofcontents

\section{Introduction}

The analysis of perturbed black holes requires, at a minimum, understanding the linearized field equations, i.e., the linearized Einstein equations (EEs) in the case of gravitational perturbations, $h_{ab}$. Due to the diffeomorphism covariance of the EEs, these are deterministic only after passing to an appropriate gauge such as the Lorenz gauge, $\nabla^a(h_{ab}-\frac{1}{2}hg_{ab})=0$, where $h=g^{ab}h_{ab}$. In fact, as is well known, the linearized EEs in the Lorenz gauge are strongly hyperbolic\footnote{See, e.g., \cite{Reula} for a precise definition of this notion and some of its consequences.}, and, as a consequence, have a well-posed initial value formulation. This implies, in particular, the existence of unique causal, i.e., advanced or retarded, Green's functions. Furthermore, 
singularities propagate along null geodesics as specified by the propagation-of-singularities theorem \cite{hormander1,hormander2}. For example, this theorem immediately shows that Lorenz-gauge gravitational perturbations sourced by a stress energy tensor for a particle on a timelike geodesic\footnote{I.e., $T^{ab}(x) = m\int {\rm d}\tau \delta(x,\gamma(\tau)) \dot \gamma^a \dot \gamma^b$, where $\gamma(\tau)$ is a timelike geodesic.} have no singularities off of this geodesic. 

A general practical problem with the perturbative EEs in Kerr spacetime in the Lorenz gauge -- and,  to the best of our knowledge, in any other gauge -- is that, unlike in Schwarzschild, there is no known basis of tensors in which the equation becomes separable.
The common way to circumvent this difficulty is to pass to the Teukolsky formulation \cite{Teukolsky:1972my,Teukolsky:1973ha}, wherein the perturbed Weyl scalars $\psi_0$ or $\psi_4$ of $h_{ab}$, instead of $h_{ab}$ itself, are the main objects. The resulting decoupled equation is not only scalar but also separable, and therefore, amenable to frequency domain techniques in an efficient way. It is also strongly hyperbolic, and consequently, has a well-posed initial value formulation and controlled propagation of singularities, just as the linearized EEs in Lorenz gauge. Since the Weyl scalars $\psi_0$ and $\psi_4$ furthermore encode the information about energy-momentum fluxes at the horizon and null infinity, their knowledge is entirely sufficient as long as one does not require the metric perturbation itself, as would be the case e.g. for self-force calculations (for reviews, see, e.g.,~\cite{Poisson:2011nh,Pound:Wardell}) or when going beyond linear order in perturbation theory. 

Actually, as discovered already shortly after Teukolsky's work, the metric perturbation $h_{ab}$ can be recovered from a solution, $\Phi$, to the Teukolsky equation to a certain extent. This so-called ``reconstruction procedure'' \cite{Chrzanowski:1975wv,Kegeles:1979an} works by applying a certain second-order tensor-valued differential operator ${\mathcal S}^\dagger_{ab}$ [see Eq.~\eqref{eq:Sdag}] to $\Phi$, such that $h_{ab} \equiv {\rm Re} \, \S^\dagger_{ab} \Phi$ is a solution to the homogeneous linearized EEs. Furthermore, it is known that, up to a gauge transformation and an infinitesimal perturbation towards another member of the Kerr family, in a suitable sense\footnote{For mode solutions, this was argued in \cite{Ori:2002uv}. More generally, this can be shown for any solution having suitable Bondi-type decay at $\mathscr{I}$; see \cite{HT1}.} every solution to the \emph{homogeneous} EEs can be written in this form for a suitable ``Hertz potential'', $\Phi$. The Hertz potential is related to $\psi_0$ by a simple, well-known formula [see Eq.~\eqref{eq:th4}].

By construction, a metric perturbation $h_{ab} = {\rm Re}\, \S^\dagger_{ab} \Phi$ in reconstructed form is in a so-called ``traceless ingoing radiation gauge'' (TIRG), meaning that $l^a h_{ab} = 0 = h$, where $l^a$ is the outgoing principal null direction. As appreciated clearly in \cite{Price,Price2,Ori:2002uv}, this means that, purely for algebraic reasons, it is impossible to write a solution to the \emph{inhomogeneous} EEs in reconstructed form unless the source $T_{ab}$ in that equation has the non-generic and undesirable property that $T_{ab}l^b=0$. For similar reasons, it is impossible for there to exist causal Green's functions in TIRG.

As shown by \cite{Green:2019nam}, in the following referred to as GHZ (after the authors Green, Hollands, and Zimmerman), the situation is actually not as bad as it may seem, since one can save the metric reconstruction procedure if one is prepared to give up the trace-free ($h=0$) part of the TIRG condition (though not $h_{ab}l^b=0$), and if one is willing to add a certain correction piece, $x_{ab}$, to ${\rm Re} \, \S^\dagger_{ab} \Phi$. The correction piece is determined by a set of transport equations \cite{Green:2019nam}, which can be solved by integration. In the present work, we shall use these insights to construct causal Green's functions for the ingoing radiation gauge (IRG) -- meaning in this paper that $h_{ab}l^b=0$ but not necessarily $h=0$. These Green's functions have two pieces. The first is a \emph{reconstructed piece}, involving a causal solution to a sourced Teukolsky equation. That piece has been constructed already by Ori \cite{Ori:2002uv} using frequency domain techniques. In \cite{Ori:2002uv}, it was clearly appreciated that the reconstructed piece alone falls short of defining a proper solution to the sourced EEs. In this paper, we show how to remedy this by adding a \emph{correction piece} corresponding to $x_{ab}$. Thus, we will show (see theorem \ref{thm:prop}) that the full retarded $(+)$ and advanced $(-)$ IRG Green's functions to the linearized EEs take the general form 
   \begin{equation}
   \label{PropHX}
       {}^\pm \mathcal{G}_{ab}{}^{a'b'}(y,y') =
       {\rm Re} \bigg[ {\mathcal S}^\dagger_{ab} \, {}^\pm \overline{\mathcal{H}}(y,y') \, 
       \overline{\mathcal{S}}^{a'b'} \bigg] +
       {}^\pm \mathcal{X}_{ab}{}^{a' b'}(y,y'),
   \end{equation}
   where an overbar on a quantity denotes complex conjugation.
The reconstructed piece ${}^\pm \mathcal{H}$ is obtainable from specific mode solutions to Teukolsky's equation \cite{Ori:2002uv}, whereas the correction piece ${}^\pm \mathcal{X}_{ab}{}^{a'b'}$ is constructed from the GHZ transport equations. It is this correction piece which is the essential innovation of this work. By contrast to the reconstructed piece, the corrector piece is most naturally expressed in the time domain\footnote{In principle, it is straightforward to obtain a frequency domain representation if desired as well; see section \ref{sec:freqdom}.}. In section \ref{sec:Xprop}, we give a lengthy but completely explicit expression for the correction piece ${}^\pm \mathcal{X}_{ab}{}^{a'b'}$ involving step functions and rational functions of various optical scalars in Kerr, which in turn have well-known and explicit coordinate expressions. The detailed properties of our IRG Green's functions are listed in theorem \ref{thm:prop}. In particular, they satisfy the usual Green's function identity on the physically relevant subspace of \emph{conserved} stress energy sources\footnote{Note that this is different from the situation in Lorenz gauge, see remark 4 below theorem \ref{thm:prop} for further explanations.}. They also have a causal support by construction. 

It is easy to see that the linearized EEs in IRG fail to be strongly hyperbolic in the sense of \cite{Reula} because Cauchy surfaces are not even non-characteristic 
for the EEs in this gauge. This is closely related to the fact that some of the EEs in IRG are in fact transport equations. Thus, the existence of causal Green's functions does not follow from general principles alone as in the case of the Lorenz gauge, but is a consequence of more detailed features of the EEs on Kerr. As a consequence, the  propagation-of-singularities property as familiar from Lorenz gauge is no longer guaranteed on general grounds, and in fact fails in IRG. We  give a precise comparison between the propagation of singularities in Lorenz gauge and IRG in terms of wave front sets in section \ref{sec:propsing}.

The difference between the two gauges in this regard is  illustrated, e.g., by the well-known \cite{Pound:2013faa} fact that the IRG suffers from certain string-like gauge singularities. For example, by contrast to the Lorenz gauge, the singularity of the stress energy tensor of a point-like particle on a time-like geodesic would be propagated along the outgoing principal null geodesics emanating from the source ``backward'' (for $-$) or ``forward'' (for $+$). Such gauge singularities obviously could be removed by passing to an alternative gauge such as the no-string gauge, or to a Lorenz gauge; see \cite{HT2} respectively \cite{Green2,Dolan,Dolan2}, where prescriptions are given  for constructing the prerequisite gauge vector field in each of these approaches.

Our IRG Green's functions are potentially useful for higher-order gravitational self-force (GSF) calculations in Kerr, which require the second-order metric perturbation sourced by the first-order perturbation; see, e.g., \cite{Pound:2012dk,Pound:Wardell,Spiers:2023cip,Spiers:2023mor}. In fact, our Green's functions reduce the labor of finding this second-order metric perturbation, because, for example, the integrations required to find the completion piece encoded in ${}^\pm \mathcal{X}_{ab}{}^{a'b'}$ are, by our explicit formulas in section \ref{sec:Xprop}, reduced to a single integration arising when computing the convolution of \eqref{PropHX} with the source. Obtaining the GSF itself would further require subtracting an appropriately chosen singular part. It would be interesting to construct this singular subtraction piece directly in IRG by performing a short distance analysis of \eqref{PropHX}. We leave this interesting problem for future work.

Our Green's functions may also be useful for quantizing the metric perturbation directly in IRG  based on the quantization of the spin-2 Hertz-potentials and Weyl-scalars discussed recently in \cite{Iuliano:2023bqp}. For example, in the usual quantization approaches to theories without local gauge invariance such as the free Klein-Gordon quantum field theory, the retarded minus advanced Green's function express the spacetime covariant commutator of the quantized perturbation in IRG in a canonical quanitzation, see e.g., \cite{Hollands:2014eia}. One might therefore expect that our Green's functions will appear in the commutator function for the quantum metric perturbation IRG, if indeed this could be defined.

{\bf Conventions:} We generally follow the conventions of 
\cite{Waldbook}, except for the metric signature, which we take as 
$(+,-,-,-)$ in accordance with most of the literature about the Geroch-Held-Penrose (GHP) formalism. We also take $8\pi G=c=1$. Linear operators are denoted by calligraphic letters.

\section{Black hole perturbation theory}

\subsection{Kerr metric, NP frames, and GHP formalism}

We consider the part of  Kerr spacetime corresponding to the exterior of the black hole, between the future and past outer horizons $\sH^\pm$ and the future and past null-infinities $\sI^\pm$, 
in the subextremal parameter range $0 \le M < |a|$. 
The Kerr metric is a type D spacetime with two repeated principal null directions called $n^a$ and $l^a$. In Boyer-Lindquist (BL) coordinates $(x^\mu) = (t,r,\theta,\varphi)$, they are given in the Kinnersley normalization by 
\begin{subequations}\label{eq:Kintet}
  \begin{align}
    (l^\mu) &= \frac{1}{\Delta} \Big( r^2+a^2, \Delta, 0, a \Big), \\
    (n^\mu) &= \frac{1}{2 \Sigma} \Big( r^2+a^2, -\Delta,0,a \Big), \\
    (m^\mu) &= \frac{1}{\sqrt{2} (r+ia\cos\theta)} \Big( ia\sin\theta, 0,1,i\csc\theta \Big),
  \end{align}
\end{subequations}
where $\Delta = r^2 -2Mr+a^2$ and $\Sigma = r^2 + a^2 \cos^2 \theta$. The 
greater and smaller roots of $\Delta$ (outer and inner horizon radii) are denoted by $r_\pm$, so that the exterior region corresponds to $r>r_+$.
The vector $m^a$ completes $n^a, l^a$ to a complex null (Newman-Penrose, NP) tetrad.
The Kinnersley tetrad \eqref{eq:Kintet} is regular at $\sH^-$
and well-behaved at $\sI^+$, and for this reason is particularly  useful when considering expressions related to the IRG.
The Kerr metric itself is
\begin{equation}
g^{ab} = 2(l^{(a}n^{b)}-\bar m^{(a} m^{b)}),
\end{equation}
corresponding to the normalizations $l^a n_b = 1, m^a \bar m_a = -1$, with all other contractions equal to zero.

Advanced and retarded Kerr-Newman (KN) coordinates are defined by, respectively,
\begin{equation}
\label{eq:KNdef}
(x^\mu) = (v=t+r_*(r), r, \theta, \varphi^*), \quad 
(x^\mu) = (u=t-r_*(r), r, \theta, \varphi_*),
\end{equation}
where
\begin{subequations}
\begin{align}
\dd \varphi^* =& \ \dd \varphi + a \frac{\dd r}{\Delta}, \\
\dd \varphi_* =& \  \dd \varphi - a \frac{\dd r}{\Delta},
\end{align}
\end{subequations}
and where 
\begin{equation}
\dd r_*=\frac{r^2+a^2}{\Delta}\dd r\, 
\end{equation}
defines the Kerr tortoise coordinate $r_*$.
The Kinnersley tetrad \eqref{eq:Kintet} in retarded KN coordinates $(x^\mu)=(u,r,\theta,\varphi_*)$ is given by:
\begin{subequations}
\label{eq:Kintet up}
\begin{align}
(l^\mu)&=\Big( 0,1,0,0 \Big), \\
(n^\mu)&=\frac{1}{\Sigma} \Big(r^2+a^2,-\Delta/2,0,a \Big) ,\\
(m^\mu) &=\frac{1}{\sqrt{2}(r+ia\cos\theta)}\Big( ia \sin\theta,0,1, i\csc \theta \Big).
\end{align}
\end{subequations}
An NP frame aligned with the two principal null directions is
 unique only up to a local boost of $n^a$, $l^a$ and a local rotation of
 $m^a$, $\bar{m}^a$.
In order to keep track of this dependence in a 
transparent way, we will employ the Geroch-Held-Penrose (GHP) formalism~\cite{GHP} in this paper. In the GHP formalism, it is trivial to transform any geometric quantity or equation from one frame to another -- while keeping $l^a$ and $n^a$ aligned along the principal null directions -- or from one coordinate system to another. Nevertheless, it can sometimes be useful to consider  specific NP tetrads and coordinates, e.g., when taking limits to $\sI^+$ or $\sH^-$
in IRG. The Kinnersley frame is regular at $\sI^+$ and $\sH^-$, and is therefore useful in this regard. Likewise, when evaluating expressions in this paper
related to the outgoing radiation gauge (ORG; by which we mean $h_{ab}n^b=0$ in this paper),  or when taking limits to $\sI^-$ or $\sH^+$, advanced KN coordinates would be more natural, as well as a frame obtained from that above by 
the $t-\varphi$-reflection isometry of Kerr.

In the GHP formalism~\cite{GHP}, 
one considers the Weyl components,
\begin{subequations}\label{eq:Weylcomp}
\begin{align}
\Psi_0  = -C_{abcd} \; l^a m^b l^c m^d, \\
\Psi_1  = - C_{abcd} l^a n^b l^c m^d, \\
\Psi_2  = -\half C_{abcd}( l^a n^b l^c n^d + l^a n^bm^c \bar{m}^d),\\
\Psi_3  = -C_{abcd} l^a n^b \bar{m}^c n^d, \\
\Psi_4  = -C_{abcd} \; n^a \bar{m}^b n^c \bar{m}^d,
\end{align} 
\end{subequations}
and the optical scalars 
\begin{subequations}\label{GHPscalars}
\begin{align}
\kappa &=m^a l^b \nabla_b l_a \circeq \GHPw{3}{1} ,\\
\tau &= m^a n^b \nabla_b l_a \circeq \GHPw{1}{-1},\\
\sigma &= m^a m^b \nabla_b l_a \circeq \GHPw{3}{-1}, \\
\rho &= m^a  \bar{m}^b \nabla_b l_a  \circeq \GHPw{1}{1},
\end{align}
\end{subequations}
together with their primed counterparts $\kappa',\tau',\sigma',\rho'$ defined by exchanging $l^a \leftrightarrow n^a, m^a \leftrightarrow \bar{m}^a$. 
The notation $X \circeq \GHPw{p}{q}$ indicates that a quantity $X$ transforms as a GHP scalar with the scaling weights $(p,q)$ under the remaining tetrad rescalings \cite{GHP}. In our NP tetrad aligned with the principal null directions, 
we have $\Psi_i = 0, i \neq 2$ and $\kappa=\sigma=\kappa'=\sigma'=0$. See, e.g., \cite{Price} for the values of the remaining GHP quantities in various frames.

For GHP calculus we 
require a GHP covariant derivative, denoted by $\Theta_a$. Its action on a GHP scalar $\eta$ with scaling weights 
$\eta \circeq \GHPw{p}{q}$ is 
\begin{align}\label{nabladef}
\Theta_a \eta &= \bigg[\nabla_a - \half(p+q) n^b \nabla_a l_b + \half(p-q) \bar{m}^b \nabla_a m_b\bigg]\eta \nonumber \\
 & \equiv \bigg[\nabla_a + l_a(p\epsilon'+q\bar{\epsilon}') + n_a(-p\epsilon-q\bar{\epsilon}) - m_a(p \beta' -q \bar{\beta}) - \bar{m}_a(-p \beta +q \bar{\beta}')\bigg]\eta. 
\end{align}
In other words, the action of the GHP covariant derivative $\Theta_a$ on a GHP quantity with homogeneous transformation law gives another such quantity. 
The GHP-covariant directional derivatives along the tetrad legs are denoted traditionally by
\begin{equation}
\thorn = l^a \Theta_a, \quad 
\thorn' = n^a \Theta_a, \quad 
\eth = m^a \Theta_a, \quad 
\eth' = \bar{m}^a \Theta_a . 
\end{equation}
These operators shift the GHP weights by the amounts $\thorn: \GHPw{1}{1}, \thorn': \GHPw{-1}{-1}, \eth: \GHPw{1}{-1}, \eth': \GHPw{-1}{1}$.
In order to write these operators explicitly, one requires the remaining 4 complex spin coefficients $\epsilon$, $\epsilon'$, $\beta$, $\beta'$, which 
may be read off from \eqref{nabladef}, and which are given by
\begin{subequations}\label{GHPscalars1}
\begin{align}
\beta &=\half ( m^a n^b \nabla_a l_b - m^a \bar{m}^b \nabla_a m_b)  ,\\
\epsilon &= \half ( l^a n^b \nabla_a l_b - l^a \bar{m}^b \nabla_a m_b) ,
\end{align}
\end{subequations}
together with their primed counterparts. 
They do not have definite GHP weight (i.e., they are not GHP scalars), but in effect form part of the definition of the GHP covariant derivative $\Theta$ \eqref{nabladef}, or equivalently, of $\thorn$, $\thorn'$, $\eth$, $\eth'$. They may never appear explicitly in any GHP covariant equation nor in any GHP covariant calculation.

\subsection{Teukolsky equation and metric reconstruction}

Teukolsky's equations \cite{Teukolsky:1973ha,Teukolsky:1972my} are traditionally given in terms of the perturbed Weyl scalars $\psi_0 \equiv \delta \Psi_0$
and $\psi_4 \equiv \delta \Psi_4$ as $\O \psi_0 = {}_{+2}T, \O' \psi_4 = {}_{-2}T$, where the Teukolsky sources, ${}_{+2}T$ and ${}_{-2}T$, are constructed out of second derivatives 
of the stress tensor $T_{ab}$ in the linearized EEs $(\E h)_{ab} = T_{ab}$. Here $\E$ is the linearized Einstein operator, 
\begin{align}\label{eq:linearE}
\delta G_{ab} = (\mathcal{E} h)_{ab} \equiv \frac{1}{2} \big[ & -\nabla^c\nabla_c h_{ab} - \nabla_a\nabla_bh + 2 \nabla^c\nabla_{(a} h_{b)c} \nonumber\\
 & + g_{ab}(\nabla^c \nabla_c h - \nabla^c\nabla^d h_{cd}) \big],
\end{align}
and $\O$ and $\O'$ are the spin $s=+2$ and $s=-2$ Teukolsky operators, respectively. 

In order to make contact with the metric reconstruction
procedure by \cite{Chrzanowski:1975wv,Kegeles:1979an}, and following \cite{Wald}, we view the existence of the Teukolsky equation as a manifestation of an 
operator identity:
\begin{equation}\label{eq:SE=OT}
\S \E = \O \T,
\end{equation}
where $\S$ is the operator that prepares the ${}_{+2}T$ Teukolsky source from $T_{ab}$, and $\T$ is the operator that prepares $\psi_0$ 
from $h_{ab}$. Concretely, 
\begin{equation}\mathcal{T}h = - \frac{1}{2}Z^{bcda} \nabla_a \nabla_b h_{cd},
\end{equation}
where $Z^{abcd} = Z^{ab}Z^{cd}, Z = l \wedge m$,  
\begin{equation}
\mathcal{O} \eta := \left[g^{ab}(\Theta_a + 4B_a)(\Theta_b + 4 B_b) - 16 \Psi_2\right]\eta , 
\end{equation}
for any $\eta \circeq \GHPw{4}{0}$, and 
where $B_a = -\rho n_a + \tau \bar{m}_a \circeq \GHPw{0}{0}$, and 
\begin{equation}
\mathcal{S}T := Z^{bcda} (\nabla_a + 4 B_a) \nabla_b T_{cd}.
\end{equation}
These tensorial forms of $\mathcal{O}$, $\mathcal{S}$ and $\mathcal{T}$ appeared in~\cite{Aksteiner:2014zyp,Araneda}. For forms of 
the operators $\E$, $\S$, $\O$ and $\T$ in terms of the optical scalars and directional derivatives in the GHP formalism see e.g., \cite{Green:2019nam}. Besides
$\S \E = \O \T$ we also have its primed counterpart, its formal adjoint
\begin{equation}
\label{eq:Teukmaster}
\E \S^\dagger = \T^\dagger \O^\dagger,
\end{equation}
and its primed adjoint. Note that the adjoint of a GHP covariant operator acting on some GHP scalar $\eta \circeq \GHPw{p}{q}$ is defined 
by the usual integration by parts procedure \emph{without} any complex conjugation. The adjoint is therefore an operator on GHP
scalars of the dual weights, $\circeq \GHPw{-p}{-q}$. One has $\Theta_a^\dagger = -\Theta^a$ for instance, as well as $\E^\dagger = \E$ (used above)
and 
\begin{equation}\label{eq:Odagger}
\mathcal{O}^\dagger \eta  = \left[g^{ab}(\Theta_a - 4 B_a)(\Theta_b - 4 B_b) - 16 \Psi_2 \right] \eta 
\end{equation}
for the adjoint of Teukolsky's operator $\O$. Due to the relation $\Psi_2^{-4/3} \O' \Psi^{4/3}_2 = \O^\dagger$, $\O^\dagger$ has 
similar properties as $\O'$. Finally, one has 
\begin{align}
\label{eq:Sdag}
  (\mathcal{S}^\dagger\Phi)^{cd} = Z^{b(cd)a} (\Theta_b + B_b) (\Theta_a - 3 B_a) \Phi .
\end{align}

Equation \eqref{eq:Teukmaster} forms the basis of the reconstruction method by \cite{Chrzanowski:1975wv,Kegeles:1979an}: 
If $\Phi \circeq \GHPw{-4}{0}$ is a solution to $\O^\dagger \Phi = 0$, then $\S^\dagger_{ab} \Phi$ is a complex-valued solution 
to the homogeneous linearized EEs. Since $\E$ is a real operator, the real (or imaginary) part of $\S^\dagger_{ab} \Phi$ is also a solution 
to the homogeneous linearized EEs. 
Thus
\begin{equation}
\label{hreconstr}
h_{ab} = \Re(\S^\dagger \Phi)_{ab} = \Re \left[Z_{d(ab)c} (\Theta^d + B^d) (\Theta^c - 3 B^c) \Phi \right]
\end{equation}
solves $(\E h)_{ab}=0$ whenever $\O^\dagger \Phi = 0$. A metric of this form is said to be of ``reconstructed form''.
 In Minkowski spacetime, the reconstructed ansatz \eqref{hreconstr} for $h_{ab}$ may be connected with the commonly used basis of plane wave solutions as follows. 
We define a flat NP tetrad by $
l^a = \frac{1}{\sqrt{2}}[(\partial_0)^a + (\partial_1)^a]$, $n^a = \frac{1}{\sqrt{2}}[(\partial_0)^a - (\partial_1)^a]$, $m^a = \frac{1}{\sqrt{2}}[(\partial_2)^a + i(\partial_3)^a]$, we let $p_a$ be a constant null vector, and we let 
$\Phi(x) = (h^+ + ih^\times) \sin(px)$, which for constant amplitudes $h^{+,\times} \in {\mathbb R}$ is a solution to $\O^\dagger \Phi = 0$ in Minkowski spacetime. Then the 
reconstructed metric perturbation \eqref{hreconstr} is
\begin{equation}
    h_{ab} = \left[ h^+ \epsilon_{ab}^+(p)
    + h^\times \epsilon_{ab}^\times(p)
    \right] \sin (px)
\end{equation}
where each $\epsilon^{+,\times}_{ab}(p) \equiv \mp \Re/\Im 
(Z_{cabd} p^c p^d)$ has the properties of a polarization tensor in Minkowski spacetime \cite{Green:2022htq}.

In Kerr, one can show \cite{HT1} that under suitable global decay assumptions at $\sI^+$, \emph{all} solutions to the homogeneous EEs can be put into 
reconstructed form, up to a zero mode, $\dot g_{ab}$, meaning an infinitesimal perturbation within the Kerr family, see appendix \ref{app:D}, and a pure 
gauge perturbation, $({\mathcal L}_X g)_{ab}$. Namely, assume that in a suitable gauge the metric perturbation 
has a Bondi-type decay near $\sI^+$, i.e., we have in the Kinnersley NP frame,
\begin{equation}
\label{eq:hdec}
h_{ll}, h_{lm}, h_{ln}, h_{m\bar{m}} = O\left(\frac{1}{r^2} \right), \quad
h_{nn}, h_{nm}, h_{mm} = O\left(\frac{1}{r} \right)
\end{equation}
as $r \to \infty$ at fixed retarded KN coordinates $(u,\varphi_*,\theta)$. Furthermore, assume that -- possibly after subtracting a zero mode -- 
all the coefficients of \eqref{eq:hdec} in a $1/r$ expansion have a 
square-integrable decay as $u \to -\infty$. Then, if $(\E h)_{ab} = 0$, there exists a 
``Hertz potential'' $\Phi \circeq (-4,0)$ solving $\O^\dagger \Phi = 0$, 
a zero mode $\dot g_{ab}$, and a gauge vector field $X^a$ such that 
\begin{equation}
\label{hab:rec}
h_{ab} = \Re(\S^\dagger \Phi)_{ab} + \dot g_{ab} + (\mathcal{L}_X g)_{ab}, 
\end{equation} 
and such that $\Phi = O(r^3)$ as $r \to \infty$ in retarded KN coordinates. For a previous result
valid for individual modes, see \cite{Ori:2002uv}, and for results stating a representation of the above 
form in an approximate sense, see \cite{Green:2019nam,Prabhu:2018jvy}. 

\section{The GHZ method}
\label{ch:GHZ}

\subsection{GHZ decomposition}
\label{sec:GHZ decomposition}

Now consider the case when $h_{ab}$ satisfies the \emph{inhomogeneous} linearized EEs off of the Kerr geometry with a given stress-energy tensor $T_{ab}$,
\begin{equation}\label{eq:Eh=T}
(\E h)_{ab}=T_{ab}.
\end{equation}
Unlike for the case $T_{ab}=0$, it is in general no longer possible 
to write $h_{ab}$ in reconstructed form \eqref{hab:rec}. This can 
be seen as follows \cite{Price}. Neither $\dot g_{ab}$ nor $({\mathcal L}_X g)_{ab}$ 
can give a non-trivial contribution to $T_{ab}$ -- the linearized Einstein operator, $\E$, annihilates any pure gauge perturbation and any zero mode.
Also, a reconstructed perturbation, $h_{ab} = \Re(\S^\dagger \Phi)_{ab}$, automatically is in the {\it traceless ingoing radiation gauge} (TIRG), meaning that\footnote{In this paper, as alluded to in the introduction, we will use the term IRG for any perturbation satisfying the first three conditions, $h_{lm} = h_{ln} = h_{ll} = 0$, but possibly not $h_{m\bar{m}} = 0$.}, 
\begin{equation}\label{IRG}
h_{lm} = h_{ln} = h_{ll} = h_{m\bar{m}} = 0.
\end{equation}
Inspection of the $ll$-component of the linearized Einstein operator in GHP form given, e.g., in the appendix of \cite{Green:2019nam} then immediately shows that, 
automatically, $(\E h)_{ll} = 0$. Thus, if $(\E h)_{ll} \neq 0$, or said differently, 
if $T_{ll} \neq 0$, then it is simply impossible to write $h_{ab}$ in reconstructed form up to a gauge perturbation or zero mode. 

Actually, one may suspect that the situation is not as bad as it may seem, because, except for $h_{m\bar{m}}=0$, i.e., for the traceless condition $h=0$, all conditions for the TIRG \eqref{IRG} may be imposed by adding, if necessary, a pure gauge perturbation $(\mathcal{L}_X g)_{ab}$ to $h_{ab}$ in an algebraically special background \cite{Green:2019nam,Andersson:2021eqc}. However, we have to accept that the remaining gauge condition $h_{m\bar{m}}=0$ simply may not be imposed in general. Taking as a hypothesis that the $l$-components of the stress energy tensor $T_{ab}$ in equation \eqref{eq:Eh=T} are the essential obstacle against writing $h_{ab}$ in reconstructed form -- thus in particular in TIRG --, Ref.~\cite{Green:2019nam} proposed to subtract from $h_{ab}$ a term called a corrector, $x_{ab}$, which in effect achieves $T_{ab}l^a = 0$. Said differently, we seek an $x_{ab}$ such that
\begin{equation}\label{xeq}
(\mathcal{E} x)_{ab} l^b = (\E h)_{ab} l^b
\end{equation}
so that $h_{ab} - x_{ab}$ satisfies the linearized EEs with a new source $S_{ab} = T_{ab} - (\mathcal{E} x)_{ab}$ such that $S_{ab}l^b = 0$. 

Of course we could choose $x_{ab}$ to be $h_{ab}$ itself, but this would obviously not bring about any simplification. Instead, we want $x_{ab}$ to satisfy an equation which is simpler than the original linearized sourced EEs -- hoping that we may then subsequently find $h_{ab} - x_{ab}$ in the form of a reconstructed metric solving a suitable sourced Teukolsky equation for $\Phi$. We just argued that the modified metric perturbation $h_{ab} - x_{ab}$ should not have a non-trivial $l$-component, so given that we may easily impose $h_{ab}l^b= 0$ by a change of gauge, we should impose $x_{ab}l^b = 0$, too. This restricts the type of NP components that $x_{ab}$ can have. 

As shown by \cite{Green:2019nam}, if we choose $x_{ab}$ to have only $nn, nm, m\bar{m}$ components, then all three NP components of equation \eqref{xeq} become {\em ordinary} differential equations for these components.  More precisely, let us consider a symmetric real tensor $x_{ab}$ with the NP components
\begin{equation}
\label{eq:xdef}
x_{ab} = 2m_{(a} \bar{m}_{b)} x_{m\bar{m}} -2l_{(a} \bar{m}_{b)} x_{nm} -2l_{(a} m_{b)} x_{n\bar{m}} + l_a l_b x_{nn}.
\end{equation}
Then the NP components of equation \eqref{xeq} are
schematically of the form
\begin{equation}\label{eq:xmmb}
\begin{split}
\rho^2 \thorn \left[ \frac{\bar{\rho}}{\rho^3} \thorn \left(\frac{\rho}{\bar{\rho}} x_{m\bar{m}} \right) \right]&= T_{ll}
\end{split}
\end{equation}
for $x_{m\bar{m}}$,
\begin{equation}
\label{eq:xnm}
\frac{\rho}{2(\rho+\bar\rho)} \thorn
\left\{ (\rho+\bar{\rho})^2
\thorn \left[ \frac{1}{\rho(\rho+\bar\rho)} x_{nm} \right] \right\} = T_{lm}
\text{$+$ (terms involving $x_{m\bar{m}}$)}
\end{equation}
for $x_{nm}$, and
\begin{equation}
\label{eq:xnn}
\half (\rho+\bar{\rho})^2\thorn \left( \frac{1}{\rho+\bar\rho}x_{nn} \right) = T_{ln}
\text{$+$ (terms involving $x_{m\bar{m}},x_{nm}$)}
\end{equation}
for $x_{nn}$. Note that this system for $x_{nn}$, $x_{nm}$ and $x_{m\bar{m}}$ is ``triangular'' in the sense that the integration of the equation for  $x_{m\bar{m}}$ does not involve  either $x_{nn}, x_{nm}$, and the integration for $x_{nm}$ does not involve $x_{nn}$. The precise form of these equations \cite{Green:2019nam}
is recalled in Eqs.~\eqref{eq:xmmb1}, \eqref{eq:xnm1} and \eqref{eq:xnn1} below for convenience.

We call these equations for the NP components $x_{nn}$, $x_{nm}$ and $x_{m\bar{m}}$ of the corrector $x_{ab}$ the {\it GHZ transport equations}, because the GHP operator $\thorn$ is simply $\partial_r$ in retarded KN coordinates $(x^\mu)=(u,r,\theta,\varphi_*)$ [see Eq.~\eqref{eq:KNdef}], and the Kinnersley frame [see Eq.~\eqref{eq:Kintet up}].
Therefore, solving the GHZ transport equations  involves $r$-integrations at constant $(u,\varphi_*,\theta)$. 

Once the corrector $x_{ab}$ has been determined, it is possible to see \cite{Green:2019nam} that the remaining part of the metric, that is $h_{ab}-x_{ab}$, can be written in reconstructed form, plus a  gauge perturbation  and an algebraically special perturbation, provided that the metric $h_{ab}$ (hence also the stress tensor $T_{ab}$) satisfies suitable fall-off conditions towards $\sI^+$ (i.e. as $r \to \infty$), or suitable regularity conditions at $\sH^-$. 

By analyzing these steps in detail, one can prove the following decomposition~\cite{Green:2019nam} (see \cite{HT1} for a proof using weaker global conditions on $h_{ab}$
and $T_{ab}$ than in \cite{Green:2019nam}):

\medskip
\noindent
{\bf GHZ decomposition}: Given $h_{ab}$ such that the stress tensor $T_{ab}$ satisfies suitable regularity conditions at $\sI^+$ respectively $\sH^-$ (detailed below in section \ref{GHZproof}), there exists a gauge vector field $X^a$, a zero mode $\dot g_{ab}$, and a corrector $x_{ab}$ (satisfying the above transport equations), such that
\begin{equation}\label{eq:decompi}
h_{ab} = \dot g_{ab} + \Re(\S^\dagger \Phi)_{ab} + x_{ab} + ({\mathcal L}_X g)_{ab}.
\end{equation}
Calculating the $\psi_0$ perturbed Weyl NP component of this perturbation and using a Teukolsky-Starobinsky identity (see e.g. \cite{Chandrasekhar:1984siy}), one sees that the 
Hertz potential $\Phi \circeq \{-4,0\}$ satisfies 
\begin{equation}\label{eq:th4}
\thorn^4 \Phi = -4 \bar{\psi}_0, 
\end{equation}
which can also be considered to be a transport equation for $\Phi$ given $\psi_0$, which satisfies the Teukolsky equation, 
\begin{equation}\label{eq:Teuk psi0}
\mathcal{O} \psi_0 = {}_{+2}T, 
\end{equation}
with ${}_{+2}T = \S T$ the spin $s=+2$ Teukolsky source. Since $x_{ab}$ can also be obtained by transport equations, the problem of determining $h_{ab}$ from the given $T_{ab}$, and the given boundary/initial conditions can be reduced to solving a Teukolsky equation, i.e. either for $\psi_0$ or for $\Phi$, and certain transport equations which can be integrated explicitly as we will see. 

\subsection{Boundary conditions for GHZ decomposition}
\label{GHZproof}

The GHZ decomposition is not unique for a given $h_{ab}$, i.e. the tensors $x_{ab}$, $\dot g_{ab}$, $X^a$ and the Hertz potential $\Phi$ are not unique. In fact, we now outline {\it two} explicit integration schemes for the various partial and ordinary differential equations required to obtain the pieces in the GHZ decomposition. Each produces a specific decomposition, and each depends on a specific choice of global boundary conditions: final conditions at $\sI^+$ or initial conditions at $\sH^-$; see \cite{HT1} and \cite{Green:2019nam} for details. Final conditions at $\sI^+$ are useful for an advanced solution, whereas initial conditions at $\sH^-$ are useful for a retarded solution in IRG.
For 
ORG,
one would consider $\sH^+$ and $\sI^-$ instead of $\sI^+$ and $\sH^-$. 

\medskip
\noindent
{\bf Backward integration.}
First, we will consider trivial final conditions at $\sI^+$, corresponding to a ``backward'' integration scheme for the various transport equations along the affinely parameterized null geodesics tangent to $l^a$ from $\sI^+$ to $\sH^-$. 
Constructing a corresponding GHZ decomposition requires certain decay conditions towards $\sI^+$ given by equations \eqref{eq:hdec} for the perturbation $h_{ab}$ and by 
\begin{equation}\label{Tdec}
T_{nn} = O\left(\frac{1}{r^2}\right) , \quad T_{ln}, T_{m\bar m}, T_{nm} = O\left(\frac{1}{r^3}\right), \quad
T_{lm}, T_{mm} = O\left(\frac{1}{r^4}\right), 
\quad
T_{ll} = O\left(\frac{1}{r^5}\right)
\end{equation}
for the NP components of the stress energy tensor. These conditions refer to retarded KN coordinates
$(r,u,\theta,\varphi_*)$, and together with \eqref{eq:hdec} are compatible with non-zero outgoing fluxes of energy, 
angular momentum carried by the matter $T_{ab}$ and with outgoing gravitational radiation (Bondi news) at $\sI^+$. 
Additionally, we require a suitable decay near $\sI^+$ as $u \to -\infty$, i.e. towards spatial infinity.

To construct the GHZ decomposition, one goes through the following steps. 

\begin{itemize}
\item
One performs a gauge transformation with gauge vector field $\xi^a$ determined so as to eliminate any $l$ NP-component of $h_{ab}$.

\item 
One obtains the corrector tensor $x_{ab}$ by backward integration (from $\sI^+$) of the GHZ transport equations \eqref{eq:xmmb1}, \eqref{eq:xnm1} and \eqref{eq:xnn1}. 

We determine a residual gauge transformation
$\zeta^a$ designed so that the metric takes the form \eqref{eq:decompi} for the combination $X^a = \xi^a + \zeta^a$.

\item
One determines $\Phi$ from the perturbed Weyl scalar $\psi_0$ via \eqref{eq:th4}, viewed again as a transport equation, by backward integration from $\sI^+$ with trivial final conditions.

\item
In its turn, $\psi_0$  is determined by solving
the Teukolsky equation $\O \psi_0 = {}_{+2}T$ with advanced boundary conditions.

\end{itemize}

The backward integration scheme is suitable for finding the advanced Green's function in an ingoing radiation gauge. 
In this situation, the support of $h_{ab}$ will be bounded in the past of some compact set (in the Lorenz gauge), so $h_{ab}$ will automatically vanish
in an open neighborhood of $\sI^+$, our conditions \eqref{eq:hdec} and \eqref{Tdec} are automatically satisfied, and the required GHZ decomposition 
already holds by the results of \cite{Green:2019nam} and does not require the more complicated analysis of \cite{HT1}.

\medskip
\noindent
{\bf Forward integration.}
One can also obtain a GHZ decomposition \eqref{eq:decompi} integrating the various transport equations ``forward'', i.e. 
from $\sH^-$ to $\sI^+$ along the affinely parameterized null geodesics tangent to $l^a$, imposing trivial initial conditions at $\sH^-$. 
This requires certain conditions on $h_{ab}$ at $\sH^-$. They are physically more restrictive than those in \eqref{eq:hdec} imposed at $\sI^+$ 
in the sense that they forbid stationary perturbation and outgoing fluxes of gravitational radiation and matter stress-energy-angular momentum at $\sH^-$. Denoting by $r_+$ the radius of the outer event horizon, the conditions are that $h_{ab}$ is twice continuously differentiable across $\sH^-$ and that, 
\begin{equation} 
\label{eq:hdec1}
\begin{split}
&h_{nn}, h_{nm},h_{m\bar m} = O((r-r_+)^2),  \\
&h_{mm}, h_{lm}, h_{ll}, h_{ln} = O(r-r_+) \quad \text{at $\sH^-$.}
\end{split}
\end{equation}
Some of the conditions 
can be seen \cite{HWcanonicalenergy} 
as mere gauge conditions and therefore do not contain any physical restrictions. The conditions combined, however, certainly give 
physical restrictions because they imply that \cite{HT1}
\begin{equation}
\label{Tdec1}
T_{nn} = T_{nm} = T_{nl}=\psi_4 = 0 \quad \text{at $\sH^-$,}
\end{equation}
i.e. one is imposing the absence of outgoing fluxes of 
gravitational radiation and stress-energy-angular momentum at $\sH^-$. Furthermore, they imply that \cite{HWcanonicalenergy}
\begin{equation}
\delta A = \delta \kappa = \delta \theta = 0 \quad \text{at $\sH^-$,}
\end{equation}
where $\delta A$, $\delta \kappa$ and $\delta \theta$ are the perturbed surface area, surface gravity of the Hawking Killing vector field $\chi^a = t^a + \Omega_+ \phi^a$ 
and the perturbed expansion of $n^a$ along $\sH^-$, respectively. 
Here, $\Omega_+ =a/(2Mr_+)$ is the angular velocity of the horizon.

The resulting forward integration scheme is suitable for obtaining a retarded solution to the linearized EEs in IRG. 
In this situation, the support of $h_{ab}$ will be bounded in the future of some compact set (in Lorenz gauge), so $h_{ab}$ will automatically vanish
in an open neighborhood of $\sH^-$, our conditions \eqref{eq:hdec1} are automatically satisfied, and  the required GHZ decomposition 
already holds by the results of \cite{Green:2019nam} and does not require the more complicated analysis of \cite{HT1}.

\section{Green's functions for the radiation gauges}
\label{sec:sourcefree}

We now outline how the GHZ method for the sourced linearized EEs may be used to construct advanced and retarded Green's functions in IRG. Details will be filled in in the subsequent sections. 
Since the out- and ingoing principal null vectors $l^a$ and $n^a$ in a type D spacetime are on the same footing, we can easily obtain all corresponding results for the ORG, $h_{ab}n^b=0$, by applying the GHP priming operation to all our equations and switching from retarded to advanced KN coordinates. 

The GHZ solution \eqref{eq:decompi} of the sourced linearized EEs is of the form $h_{ab}=\Re(\S^\dagger \Phi)_{ab} + x_{ab}$ plus a gauge part and a zero mode. For a solution in the radiation gauge, we are discarding by assumption the gauge part. In this section, we consider a retarded respectively advanced solution $h_{ab}$ 
associated with a stress tensor $T_{ab}$ supported in the causal future respectively past of a compact set\footnote{See \cite{HT2} for a construction of a large class of such tensors
using a cutoff procedure that preserves the conservation condition $\nabla^a T_{ab}=0$ in type D spacetimes.}. 
The problem that we will solve is how to obtain this $h_{ab}$ by means of a suitably constructed retarded respectively advanced Green's function. This Green's function is found by finding a Green's function for each step in the GHZ construction \eqref{eq:decompi} corresponding to the backward- (for the retarded Green's function) respectively forward (for the advanced Green's function) integration scheme. 
 
For a retarded respectively advanced solution, we start in a gauge (e.g. Lorenz gauge) where the support of $h_{ab}$ is contained in the causal future respectively past 
of the support of $T_{ab}$, ${\rm supp}(h_{ab}) \subset J^\pm({\rm supp}(T_{ab}))$. In particular, the decay assumptions \eqref{eq:hdec} and \eqref{Tdec} required for the backward GHZ scheme are satisfied at $\sI^+$ for the advanced solution, whereas the assumptions \eqref{eq:hdec1} and \eqref{Tdec1} required for the forward GHZ scheme are satisfied at $\sH^-$. Furthermore, the intersection of the support of $h_{ab}$ with a Cauchy surface does not extend to spatial infinity. Then, for example, the linearized mass $\dot M$ and angular momentum $\dot J$ vanish, and consequently the zero mode part of $h_{ab}$ vanishes. Furthermore, 
we do not need for the purposes of this paper the pure gauge part $({\mathcal L}_X g)_{ab}$ because we are not interested in the precise gauge transformation relating, say, the Lorenz gauge with the radiation gauge considered here. We may therefore assume that $h_{ab}=\Re(\S^\dagger \Phi)_{ab} + x_{ab}$ without the additional pieces in the GHZ decomposition. 

Now, the corrector $x_{ab}$ is obtained from $T_{ab}l^b$ by solving the 
transport equations \eqref{eq:xmmb}, \eqref{eq:xnm} and \eqref{eq:xnn}, which are the components of the equation
\begin{equation}\label{eq:corr eq}
(\E x)_{ab}l^b = T_{ab}l^b. 
\end{equation}
For the retarded solution, we require trivial initial data for these solutions for sufficiently large negative parameter along the null geodesics tangent to $l^a$, whereas for the advanced solution, we require trivial final data for sufficiently large positive parameter along the null geodesics tangent to $l^a$. We seek an appropriate Green's function ${}^\pm {\mathcal X}_{ab}{}^{a_{\textrm{p}} b_{\textrm{p}}}(y,y_{\textrm{p}})$ for  
the equation $(\E x)_{ab}l^b = T_{ab}l^b$, i.e. the transport equations for the retarded respectively advanced correctors $x_{ab}^\pm$, such that 
\begin{equation}
\label{propa:1}
    x_{ab}^\pm(y) 
    = \int {}^\pm {\mathcal X}_{ab}{}^{a_{\textrm{p}} b_{\textrm{p}}}(y,y_{\textrm{p}}) T_{a_{\textrm{p}} b_{\textrm{p}}}(y_{\textrm{p}}) \, \intd V_{\textrm{p}}
    \equiv {}^\pm {\mathcal X}_{ab}{}^{a_{\textrm{p}} b_{\textrm{p}}} * T_{a_{\textrm{p}} b_{\textrm{p}}}(y).
\end{equation}
The distributional kernel ${}^\pm {\mathcal X}_{ab}{}^{a_{\textrm{p}} b_{\textrm{p}}}$ to be found
is a bi-tensor which is going to contain Dirac delta-functions of the type
$\delta(u-u_{\textrm{p}}) \delta(\varphi_* - \varphi_{*\textrm{p}}) \delta(\theta-\theta_{\textrm{p}})$ that effectively restrict the 4-dimensional integration 
$\int \intd V_{\textrm{p}} \cdots = \int  \intd^4 y_{\textrm{p}} \, \sqrt{-g_{\textrm{p}}} \cdots$ to a null generator, parameterized by the retarded KN coordinate $r_{\textrm{p}} \le \, (\ge) \, r$ along  the null geodesic tangent to $l^a$ at fixed $(u,\varphi_*,\theta)$. The subscript $\pm$ indicates retarded $(+)$ or advanced $(-)$ boundary conditions, while the label `$\textrm{p}$' will from now on indicate the second point in the kernel. 

Next, according to  \eqref{eq:th4}, the reconstructed part $\Re(\S^\dagger \Phi)_{ab}$ is obtained as follows. First we form 
the source $\S T$ in Teukolsky's equation for $\psi_0$. Then we obtain $\psi_0$ by means of an advanced or retarded Green's function ${}^\pm {\mathcal G}$ for Teukolsky's operator $\O$ associated with $\O \psi_0 = \S T$. By \eqref{eq:th4}, $\psi_0$ is related to the Hertz potential $\Phi$ through $-4\psi_0 = \th^4 \bar \Phi$. Composing 
${}^\pm {\mathcal G}$ with an advanced or retarded Green's function for the GHP operator $-\tfrac{1}{4}\th^4$ ($=-\tfrac{1}{4}\partial_r^4$ in retarded KN coordinates), one can obtain a new advanced or retarded Green's function ${}^\pm {\mathcal H}$, and one can represent the advanced and retarded Hertz potentials by a convolution 
\begin{equation}
\label{propa:2}
\Phi^\pm(y) =  {}^\pm \bar {\mathcal H}* \bar \S^{a_{\textrm{p}} b_{\textrm{p}}} T_{a_{\textrm{p}} b_{\textrm{p}}} (y)
= 
    \int {}^\pm \bar {\mathcal H}(y,y_{\textrm{p}}) \, \bar \S^{a_{\textrm{p}} b_{\textrm{p}}} T_{a_{\textrm{p}} b_{\textrm{p}}} (y_{\textrm{p}}) \, \intd V_{\textrm{p}}. 
\end{equation}
Combining equations \eqref{propa:1} and \eqref{propa:2}, we are going to obtain the advanced and retarded Green's functions in IRG
in the next sections: To construct the Green's function for the GHZ transport equations, we first recall the 
explicit solution to their {\it homogeneous} counterpart, i.e. $T_{ab}=0$ in subsection \ref{sec:homGHZ}.
The Green's function ${}^\pm {\mathcal X}_{ab}{}^{a_{\textrm{p}} b_{\textrm{p}}}$ will then be deduced from this result in subsection 
\ref{sec:Xprop}. The form of the Green's function ${}^\pm {\mathcal H}$ will be deduced in subsection \ref{sec:Phiprop}. The full Green's functions for the radiation gauges will be obtained by combining ${}^\pm {\mathcal X}_{ab}{}^{a_{\textrm{p}} b_{\textrm{p}}}$ with ${}^\pm {\mathcal H}$ and the Teukolsky source- and reconstruction operators, $\S$ and $\S^\dagger$; see theorem \ref{thm:prop}.

\subsection{Solution to the homogeneous GHZ transport equations}
\label{sec:homGHZ}

To start with, we consider the general solutions to the {\it homogeneous} transport equations for the NP components of the corrector $x_{ab}$, i.e. versions of \eqref{eq:xmmb}, \eqref{eq:xnm} and \eqref{eq:xnn} with 
$T_{ab}l^b = 0$. These can be given as explicit rational functions in $\rho$ and $\bar{\rho}$ with coefficients that are annihilated by $\thorn$  \cite{HT2}. 
These calculations rely heavily on Held's coordinate-free integration method \cite{Held} (see appendix \ref{sec:Held}) and use the fact 
that the GHP operator $\thorn$ appearing in the GHZ transport equations is basically an $r$-derivative along outgoing null geodesics tangent to $l^a$. 
In fact, in Held's formalism, one choses an adapted basis, $\tilde{\thorn}' ,\tilde{\eth},\tilde{\eth}'$, of directional derivatives different from $\thorn',\eth,\eth'$ that are, in a suitable sense, ``$r$-independent''. Typically, all quantities are expanded out in powers  (or other functions) of the GHP scalar $\rho$ 
and quantities annihilated by $\thorn$, i.e. that are in a sense ``$r$-independent''. ($\rho$ is given by 
$\rho = -(r-ia\cos \theta)^{-1} $ in the Kinnersley frame and KN/BL coordinates.)
Quantities annihilated by $\thorn$, i.e. $\thorn q^\circ=0$, will be indicated by a degree mark `$\circ$'.
The basis of the calculus are the relations between Held's directional derivative operators, and the associated Held versions
of the optical scalars $\rho^{\circ \prime}$, $\tau^\circ$, $\Omega^\circ$ and $\Psi^\circ$, described in appendix \ref{sec:Held}.

Using such techniques, it was found in \cite{HT2} that the most general  solution to the \emph{homogeneous} (i.e., assuming $T_{ab}l^b=0$) GHZ transport equations
\eqref{eq:xmmb1}, \eqref{eq:xnm1} and \eqref{eq:xnn1} is given by\footnote{Note that there is a minor typo in Eq.~(63) in \cite{HT2} that is fixed here.}
\begin{equation}\label{xnn}
\begin{split}
x_{nn}^{\rm hom} =&\ 2 \Re \bigg[ \half \frac{\rho + \bar{\rho}}{\rho^2} \tilde{\thorn}' a^\circ + (1 + \rho \bar{\rho} \Omega^{2 \circ}) \tilde{\eth}'\tilde{\eth} a^\circ + 2 \left( \rho^2 + 2 \rho \bar{\rho} \right) \tau^\circ \bar{\tau}^\circ a^\circ\\
&+ \left( \frac{\rho^2}{\bar{\rho}} + \frac{3}{2} \rho \bar{\rho} \Omega^\circ \right) \Psi^\circ a^\circ - 2 \rho^{\prime \circ} a^\circ + \tilde{\thorn}' b^\circ + 2 \rho \bar{\rho} \bar{\tau}^\circ \tilde{\eth} b^\circ + 2 \rho^2 \bar{\rho} \tau^\circ \bar{\tau}^\circ b^\circ + \rho^2 \Psi^\circ b^\circ\\
&+ \rho^2 \tilde{\eth}' e^\circ + 2 \rho^2 \bar{\rho} \bar{\tau}^\circ e^\circ - \frac{\bar{\rho}}{\rho} \left( 4 \Omega^\circ + \frac{1}{\rho} \right) \tilde{\eth}' c^\circ + 4 \frac{\bar{\rho}}{\rho} \bar{\tau}^\circ c^\circ + \left( \rho + \bar{\rho} \right)  f^\circ \bigg],
\end{split}
\end{equation}
\begin{equation}\label{xnm}
x_{nm}^{\rm hom} = \bar{\rho} \Omega^\circ \tilde{\eth} a^\circ + \frac{\left( \rho + \bar{\rho} \right)^2}{\rho} \tau^\circ a^\circ + \bar{\rho} \tilde{\eth} b^\circ + \bar{\rho} \left( \rho + \bar{\rho} \right) \tau^\circ b^\circ + \frac{2 \rho - \bar{\rho}}{\rho^2} c^\circ + \rho \left( \rho + \bar{\rho} \right) e^\circ,
\end{equation}
\begin{equation}\label{xmmb}
x_{m\bar{m}}^{\rm hom} = a^\circ \left( \frac{\rho}{\bar{\rho}} + \frac{\bar{\rho}}{\rho} \right) + b^\circ \left( \rho + \bar{\rho} \right)
\end{equation}
in the source-free region. This gives the corresponding corrector tensor
\begin{equation}\label{eq:xdef1}
x_{ab}^{\rm hom} = 2m_{(a} \bar{m}_{b)} x_{m\bar{m}}^{\rm hom} -2l_{(a} \bar{m}_{b)} x_{nm}^{\rm hom} -2l_{(a} m_{b)} x_{n\bar{m}} + l_a l_b x_{nn}^{\rm hom};
\end{equation}
cf.  Eq.~\eqref{eq:xdef}.
The GHP scalars $a^\circ$, $b^\circ$, $c^\circ$, $f^\circ$, and $e^\circ$ 
contained in equations \eqref{xnn}, \eqref{xnm} and \eqref{xmmb} are annihilated by $\thorn$, i.e., they are independent of $r$ in KN coordinates and the Kinnersley frame. They are not determined by \eqref{eq:xmmb}, \eqref{eq:xnm} and \eqref{eq:xnn}. Their GHP weights are summarized in table \ref{XandOriSIntegrationConstantsWeights}.

\begin{table}
\label{XandOriSIntegrationConstantsWeights}%\addtolength{\tabcolsep}{3pt}
\begin{center}
\begin{tabular}{c c}
\br
Quantity & GHP weights \\
\mr
$a^\circ$ & $\GHPw{0}{0}$\\
$b^\circ$ & $\GHPw{-1}{-1}$\\
$c^\circ$ & $\GHPw{1}{-1}$\\
$e^\circ$ & $\GHPw{-2}{-4}$\\
$f^\circ$ & $\GHPw{-3}{-3}$\\
\br
\end{tabular}
\caption{GHP weights of the integration constants for the homogeneous GHZ transport equations \cite{HT2}.}
\end{center}
\end{table}

\subsection{Green's function for GHZ transport equation}
\label{sec:Xprop}

We now derive the retarded and advanced Green's functions for the GHZ transport equations \eqref{eq:xmmb1}, \eqref{eq:xnm1} and \eqref{eq:xnn1} sourced by a non-trivial stress tensor $T_{ab}$ via a suitable version of the variation of constants method.

For the retarded solution, we take a step supported for $r \ge r_\textrm{p}$:
\begin{equation}
\label{Th+}
\Theta^+_\textrm{p}(r) \equiv \left\{
\begin{matrix}
1 & r \ge r_\textrm{p}\\
0 & r < r_\textrm{p}
\end{matrix}
\right. \circeq \GHPw{0}{0},
\end{equation}
where $r$ is the radius in advanced KN coordinates $(r,u,\theta,\varphi_*)$. For given $T_a \equiv T_{ab}l^b$, we then try to satisfy the GHZ transport equations
\begin{equation}
l^b[\E(\Theta^+_\textrm{p}x^{\rm hom})]_{ab} = T_a \delta_\textrm{p}(r)
\end{equation}
sourced by a delta function\footnote{\label{footnote:dic}The expression for $\delta_\textrm{p}$ as given in equation \eqref{eq:501} is correct in the Kinnersly frame. As indicated, the last expression consistent with equation \eqref{eq:502} is $\delta_\textrm{p}(r) \equiv \Lambda(r) \delta(r-r_\textrm{p})$, where $\Lambda(r)$ is a frame-dependent function of $r$.} at $r_\textrm{p}$,
\begin{equation}
\label{eq:501}
\delta_\textrm{p}(r) \equiv \th \Theta_\textrm{p}^+(r) \quad \text{$\equiv \delta(r-r_\textrm{p})$ in Kinnersley frame},
\end{equation}
by adjusting the free GHP scalars appearing in the homogeneous solution $x_{ab}^{\textrm{hom}}$, listed in table \ref{XandOriSIntegrationConstantsWeights}. Note that the equation holds automatically except at $r=r_\textrm{p}$, because, except at those points, $\Theta^+_\textrm{p}(r)$ is constant and, by construction, $l^b[\E(x^{\rm hom})]_{ab}=0$, since these are precisely the homogeneous GHZ transport equations. Thus, $l^b[\E(\Theta^+_\textrm{p}x^{\rm hom})]_{ab}$ must be proportional to a delta function localized at $r=r_\textrm{p}$ or its derivatives, because
\begin{equation}
\label{eq:502}
\delta_\textrm{p} = \thorn \Theta^+_\textrm{p} \circeq \GHPw{1}{1},
\quad
\delta'_\textrm{p} = \thorn \delta_\textrm{p} \circeq \GHPw{2}{2}.
\end{equation}
The linearized EEs gives 
\begin{multline}
[\E(\Theta^+_\textrm{p}x^{\rm hom})]_{ll} = \left[\left( \frac{\rho^2}{\bar{\rho}} + \frac{\bar{\rho}^2}{\rho} - 3 \rho - 3 \bar{\rho} \right)  a^\circ + (\rho - \bar{\rho})^2  b^\circ \right] \delta_\textrm{p}\\
+ \left[\left( \frac{\rho}{\bar{\rho}} + \frac{\bar{\rho}}{\rho} \right) a^\circ + (\rho + \bar{\rho})  b^\circ \right] \delta'_\textrm{p},
\end{multline}
\begin{multline}
[\E(\Theta^+_\textrm{p}x^{\rm hom})]_{nl} =  2 \Re \bigg[ \half \frac{\rho + \bar{\rho}}{\rho^2} \tilde{\thorn}' a^\circ + (1 + \rho \bar{\rho} \Omega^{2 \circ}) \tilde{\eth}'\tilde{\eth} a^\circ + 2 \left( \rho^2 + 2 \rho \bar{\rho} \right) \tau^\circ \bar{\tau}^\circ a^\circ\\
+ \left( \frac{\rho^2}{\bar{\rho}} + \frac{3}{2} \rho \bar{\rho} \Omega^\circ \right) \Psi^\circ a^\circ - 2 \rho^{\prime \circ} a^\circ + \tilde{\thorn}' b^\circ + 2 \rho \bar{\rho} \bar{\tau}^\circ \tilde{\eth} b^\circ + 2 \rho^2 \bar{\rho} \tau^\circ \bar{\tau}^\circ b^\circ + \rho^2 \Psi^\circ b^\circ\\
+ \rho^2 \tilde{\eth}' e^\circ + 2 \rho^2 \bar{\rho} \bar{\tau}^\circ e^\circ - \frac{\bar{\rho}}{\rho} \left( 4 \Omega^\circ + \frac{1}{\rho} \right) \tilde{\eth}' c^\circ + 4 \frac{\bar{\rho}}{\rho} \bar{\tau}^\circ c^\circ + \left( \rho + \bar{\rho} \right) f^\circ \bigg] \delta_\textrm{p},
\end{multline}
\begin{multline}
[\E(\Theta^+_\textrm{p}x^{\rm hom})]_{ml} = \bigg[ \bar{\rho} (3\rho + \bar{\rho}) \tau^\circ a^\circ + \rho^2 \bar{\rho} (\rho + \bar{\rho}) \Omega^\circ \tau^\circ b^\circ - \frac{4 \rho^2 - 3 \rho \bar{\rho} + \bar{\rho}^2}{\rho^2} c^\circ + \rho (\rho^2 + \bar{\rho}^2) e^\circ\\
- \frac{\rho^2 + \bar{\rho}^2 - 4 \rho \bar{\rho}}{2 \rho} \tilde{\eth} a^\circ - \half \bar{\rho} (\rho - 3 \bar{\rho}) \tilde{\eth} b^\circ \bigg] \delta_\textrm{p} + \bigg[ \frac{1}{2 \rho} (\rho + \bar{\rho})^2 \tau^\circ a^\circ\\
+ \half (\rho + \bar{\rho})^2 \tau^\circ b^\circ + \frac{2 \rho - \bar{\rho}}{2 \rho^2} c^\circ + \half (2 \rho^2 + \bar{\rho}^2 - \rho \bar{\rho}) e^\circ + \half \bar{\rho} \Omega^\circ \tilde{\eth} a^\circ + \half \bar{\rho} \tilde{\eth} b^\circ \bigg] \delta'_\textrm{p}.
\end{multline}
In order to manipulate the delta function terms, we note the following identities:
\begin{equation}
\begin{split}\label{eq:delta'identity}
G(r) \delta_\textrm{p}(r) + F(r) \delta'_\textrm{p}(r) =& \left[ G(r) - \thorn F(r) \right] \delta_\textrm{p}(r) + \thorn \left[ F(r) \delta_\textrm{p}(r) \right]\\
=& \left[ G(r_\textrm{p}) - \thorn F(r_\textrm{p}) \right] \delta_\textrm{p}(r) + \thorn \left[ F(r_\textrm{p}) \delta_\textrm{p}(r) \right]\\ =& \left[ G(r_\textrm{p}) - \thorn F(r_\textrm{p}) \right] \delta_\textrm{p}(r) + F(r_\textrm{p}) \thorn \delta_\textrm{p}(r).
\end{split}
\end{equation}
By way of the identity \eqref{eq:delta'identity}, we now aim to eliminate the terms proportional to $\delta'_\textrm{p}$. This, in turn, leads us to demand
\begin{equation}\label{eq: b0 in terms of a0}
b^\circ = - \frac{\rho^2_\textrm{p} + \bar{\rho}^2_\textrm{p}}{\rho_\textrm{p} \bar{\rho}_\textrm{p} (\rho_\textrm{p} + \bar{\rho}_\textrm{p})} a^\circ,
\end{equation}
\begin{equation}\label{eq: e0 in terms of a0 and c0}
e^\circ = 2 \frac{\bar{\rho}^2_\textrm{p}}{\rho^2_\textrm{p} (\rho_\textrm{p} + \bar{\rho}_\textrm{p})^2} \tilde{\eth} a^\circ - 4 \frac{\bar{\rho}^2_\textrm{p}}{\rho_\textrm{p} (\rho_\textrm{p} + \bar{\rho}_\textrm{p})^2} \tau^\circ a^\circ - \frac{2 \rho_\textrm{p} - \bar{\rho}_\textrm{p}}{\rho^3_\textrm{p} (\rho_\textrm{p} + \bar{\rho}_\textrm{p})} c^\circ.
\end{equation}
Consequently, $[\E(\Theta^+_\textrm{p}x^{\rm hom})]_{ab}l^b$ is brought into the following form:
\begin{equation}
[\E(\Theta^+_\textrm{p}x^{\rm hom})]_{ab} l^b = T_a|_{r_\textrm{p}} \delta_\textrm{p},
\end{equation}
where $\delta_\textrm{p}(r) \equiv \Lambda(r_\textrm{p}) \delta(r-r_\textrm{p})$ in this equation (see footnote \ref{footnote:dic}), and where $T_a$ is the real co-vector with NP components
\begin{equation}\label{def:t_l in terms of a0}
T_l = - 8 \frac{\rho_\textrm{p} \bar{\rho}_\textrm{p}}{\rho_\textrm{p} + \bar{\rho}_\textrm{p}} a^\circ,
\end{equation}
\begin{equation}\label{def:t_m in terms of a0 and c0}
T_m = 4 \frac{\rho_\textrm{p} \bar{\rho}^2_\textrm{p}}{(\rho_\textrm{p} + \bar{\rho}_\textrm{p})^2} \tilde{\eth} a^\circ - 4 \frac{\rho_\textrm{p} \bar{\rho}^2_\textrm{p} (3 \rho_\textrm{p} + \bar{\rho}^2_\textrm{p})}{(\rho_\textrm{p} + \bar{\rho}_\textrm{p})^2} \tau^\circ a^\circ - 6 \frac{\rho_\textrm{p}}{\rho_\textrm{p} + \bar{\rho}_\textrm{p}} c^\circ.
\end{equation}
The NP component $T_n$ is simply $[\E(\Theta^+_{\textrm{p}}x^{\rm hom})]_{nl}$  with $b^\circ$ and $e^\circ$ substituted according to \eqref{eq: b0 in terms of a0} and \eqref{eq: e0 in terms of a0 and c0}. 

We now solve for $a^\circ$, $b^\circ$, $c^\circ$, $e^\circ$ and $f^\circ$ in terms of the NP components of $T^a|_{r_\textrm{p}} \equiv T_\textrm{p}^a$ by inverting the equations \eqref{def:t_l in terms of a0} and \eqref{def:t_m in terms of a0 and c0} %and \eqref{def:t_n in terms of a0, c0 and f0} 
and using \eqref{eq: b0 in terms of a0} and \eqref{eq: e0 in terms of a0 and c0}. This leads to
\begin{equation}
\label{acirc}
a^\circ = - \frac{1}{8} \left(\frac{1}{\rho_\textrm{p}} + \frac{1}{\bar{\rho}_\textrm{p}}\right) T_{\textrm{p} \, l},
\end{equation}
\begin{equation}
\label{bcirc}
b^\circ = \frac{1}{8} \left(\frac{1}{\rho^2_\textrm{p}} + \frac{1}{\bar{\rho}^2_\textrm{p}}\right) T_{\textrm{p} \, l},
\end{equation}
\begin{equation}
\label{ccirc}
c^\circ = - \frac{1}{12} \frac{\bar{\rho}_\textrm{p}}{\rho_\textrm{p}} \tilde{\eth} T_{\textrm{p} \, l} + \frac{1}{12} \frac{\bar{\rho}_\textrm{p}}{\rho_\textrm{p}} (3 \rho_\textrm{p} + \bar{\rho}_\textrm{p}) \tau^\circ T_{\textrm{p} \, l} - \frac{1}{6} \left(1 + \frac{\bar{\rho}_\textrm{p}}{\rho_\textrm{p}}\right) T_{\textrm{p} \, m},
\end{equation}

\begin{equation}
\label{ecirc}
e^\circ = - \frac{1}{12} \frac{\bar{\rho}_\textrm{p}}{\rho^4_\textrm{p}} \tilde{\eth} T_{\textrm{p} \, l} + \frac{1}{12} \frac{\bar{\rho}^2_\textrm{p}}{\rho^4_\textrm{p}} \tau^\circ T_{\textrm{p} \, l} + \frac{1}{6} \frac{2 \rho_\textrm{p} - \bar{\rho}_\textrm{p}}{\rho^4_\textrm{p}} T_{\textrm{p} \, m},
\end{equation}

\begin{multline}
\label{fcirc}
f^\circ = \Re \bigg[ \frac{1}{4} \frac{\rho^2_\textrm{p}}{\bar{\rho}^3_\textrm{p} (\rho_\textrm{p} + \bar{\rho}_\textrm{p})} \tilde{\eth}' \tilde{\eth} T_{\textrm{p} \, l} - \frac{1}{4} \frac{\rho^6_\textrm{p} + 2 \rho^5_\textrm{p} \bar{\rho}_\textrm{p} - 3 \rho^4_\textrm{p} \bar{\rho}^2_\textrm{p} + \rho^2_\textrm{p} \bar{\rho}^4_\textrm{p} - 2 \rho_\textrm{p} \bar{\rho}^5_\textrm{p} - \bar{\rho}^6_\textrm{p}}{\rho^3_\textrm{p} \bar{\rho}^3_\textrm{p} (\rho_\textrm{p} + \bar{\rho}_\textrm{p})} \bar{\tau}^\circ \tilde{\eth} T_{\textrm{p} \, l}\\
+ \frac{1}{16} \frac{\rho_\textrm{p} (4 \rho^2_\textrm{p} + 3 \rho_\textrm{p} \bar{\rho}_\textrm{p} - 2 \bar{\rho}^2_\textrm{p})}{\bar{\rho}^3_\textrm{p} (\rho_\textrm{p} + \bar{\rho}_\textrm{p})} \Omega^\circ \tilde{\thorn}' T_{\textrm{p} \, l} + \frac{1}{4} \frac{\bar{\rho}_\textrm{p} (3 \rho_\textrm{p} + \bar{\rho}_\textrm{p}) (2 \rho^2_\textrm{p} - \bar{\rho}^2_\textrm{p})}{\rho^3_\textrm{p} (\rho_\textrm{p} + \bar{\rho}_\textrm{p})^2} \Omega^\circ \rho^{\prime \circ} T_{\textrm{p} \, l}\\
+ \half \frac{\bar{\rho}^2_\textrm{p} (2 \rho^3_\textrm{p} - 2 \rho^2_\textrm{p} \bar{\rho}_\textrm{p} - 3 \rho_\textrm{p} \bar{\rho}^2_\textrm{p} - \bar{\rho}^3_\textrm{p})}{\rho^3_\textrm{p} (\rho_\textrm{p} + \bar{\rho}_\textrm{p})^2} \tau^\circ \bar{\tau}^\circ T_{\textrm{p} \, l}\\
+ \frac{1}{16} \frac{2 \rho^7_\textrm{p} + 6 \rho^6_\textrm{p} \bar{\rho}_\textrm{p} + \rho^5_\textrm{p} \bar{\rho}^2_\textrm{p} - 21 \rho^4_\textrm{p} \bar{\rho}^3_\textrm{p} + 3 \rho^3_\textrm{p} \bar{\rho}^4_\textrm{p} + \rho^2_\textrm{p} \bar{\rho}^5_\textrm{p} - 6 \rho_\textrm{p} \bar{\rho}^6_\textrm{p} - 2 \bar{\rho}^7_\textrm{p}}{\rho^3_\textrm{p} \bar{\rho}^3_\textrm{p} (\rho_\textrm{p} + \bar{\rho}_\textrm{p})^2} \Psi^\circ T_{\textrm{p} \, l}\\
- \half \frac{2 \rho^2_\textrm{p} - \bar{\rho}^2_\textrm{p}}{\rho^3_\textrm{p} (\rho_\textrm{p} + \bar{\rho}_\textrm{p})} \tilde{\eth}' T_{\textrm{p} \, m} - \half \frac{2 \rho^3_\textrm{p} - 2 \rho_\textrm{p} \bar{\rho}^2_\textrm{p} - \bar{\rho}^3_\textrm{p}}{\rho^3_\textrm{p} (\rho_\textrm{p} + \bar{\rho}_\textrm{p})} \bar{\tau}^\circ T_{\textrm{p} \, m} \bigg] + \frac{1}{(\rho_\textrm{p} + \bar{\rho}_\textrm{p})^2} T_{\textrm{p} \, n}.
\end{multline}
We have thereby found at this stage the corrector for a source $T_{ab}$ with arbitrarily given delta-function type $l$ NP components $T_{ab}l^b = T_{\textrm{p}a} \delta(r-r_\textrm{p})$ (these are the only NP components of $T_{ab}$ that the corrector depends on). 

Our final task is to write the result for the retarded solution $x_{ab}^+$
to the GHZ transport equations in the form of equation \eqref{propa:1},
where the distributional kernel ${}^+ {\mathcal X}_{ab}{}^{a_{\textrm{p}} b_{\textrm{p}}}$ is a suitable bitensor that we shall now define.
The desired formula for the kernel ${}^+ {\mathcal X}_{ab}{}^{a_{\textrm{p}} b_{\textrm{p}}}$
follows in principle straightforwardly by substituting \eqref{acirc}, \eqref{bcirc}, \eqref{ccirc}, \eqref{ecirc} and \eqref{fcirc} into 
the solution $\Theta^+_{\textrm{p}} x_{ab}^{\textrm{hom}}$, where the homogeneous solution was given in equations \eqref{xnn}, \eqref{xnm} and \eqref{xmmb}. Carrying out these steps involves no further analysis, in principle, but bringing the final result into a displayable form requires still a very considerable amount of algebra. To display the, still lengthy but considerably shortened, result in a reasonably compact form, we introduce some auxiliary quantities. 

First, since application of the Green's function for the corrector in effect only involves $r$- but no $u,\theta,\varphi_*$ integrations in 
outgoing KN coordinates, it is useful to introduce a corresponding 
$\delta$ function just in $u,\theta,\varphi_*$. Note that in retarded KN coordinates,
\begin{equation}
    \intd V = \sqrt{-g} \, \intd^4 x = 
    \Sigma(r,\theta) \sin \theta \, \intd r \intd u \intd \theta \intd \varphi_*, 
\end{equation}
and the undensitized 4-dimensional delta function peaked at the point $y_\textrm{p}$ is 
\begin{equation}
    \delta(y,y_\textrm{p}) \equiv \frac{\delta^4(y^\mu_{}-y_\textrm{p}^\mu)}{\sqrt{-g}_{\textrm{p}}}. 
\end{equation}
We denote by $\delta^\circ$ the part of this only involving $u,\theta,\varphi_*$, or in other words, 
\begin{equation}
\label{eq:deltacirc}
\delta^\circ_\textrm{p}(y)
= \frac{\delta(u-u_\textrm{p})\delta(\theta-\theta_\textrm{p})\delta(\varphi_*-\varphi_{*\textrm{p}})}{
\Sigma_\textrm{p} \sin \theta_\textrm{p}}. 
\end{equation}
As it is, $\delta^\circ_\textrm{p}$ has only been defined in the Kinnersley tetrad and retarded KN coordinates. However, note that it is the same as formally dividing the density $\delta(y,y_\textrm{p})$ of trivial GHP weights $\{0,0\}$ by the quantity $\delta(r-r_\textrm{p})$ having GHP weights $\GHPw{1}{1}$. Thus, it is natural to define $\delta^\circ_\textrm{p}$ in a frame-invariant way by demanding that it transforms as a GHP scalar-valued distributional density 
of GHP weights $\delta^\circ_\textrm{p} \circeq \GHPw{-1}{-1}$. 
We now define 
\begin{equation}
\label{eq:Xdef}
{}^+ {\mathcal X}_{ab}{}^{a_{\textrm{p}} b_{\textrm{p}}} :=
{}^+ X_{n\, ab} l_\textrm{p}^{a_\textrm{p}}  l_\textrm{p}^{b_\textrm{p}} + 
{}^+ X_{l\, ab} l_\textrm{p}^{(a_\textrm{p}}  n_\textrm{p}^{b_\textrm{p})}  - 2 \Re \left( 
{}^+ X_{\bar m \, ab} l_\textrm{p}^{(a_\textrm{p}}  m_\textrm{p}^{b_\textrm{p})}  
\right),
\end{equation}
wherein 
\begin{equation}
\label{propa:3}
{}^+ X^{ab}_l := 2 l^a l^b \Theta^+_\textrm{p} \left[ (\rho + \bar{\rho}) \frac{1}{(\rho_\textrm{p} + \bar{\rho}_\textrm{p})^2} \right] \delta^\circ_\textrm{p},
\end{equation}
\begin{multline}
\label{propa:4}
{}^+ X^{ab}_n := \frac{1}{4} m^{(a} \bar{m}^{b)} \Theta^+_\textrm{p} \left[ (\rho + \bar{\rho}) \left(\frac{1}{\rho^2_\textrm{p}} + \frac{1}{\bar{\rho}^2_\textrm{p}}\right) - \left(\frac{\rho}{\bar{\rho}} + \frac{\bar{\rho}}{\rho} \right) \left(\frac{1}{\rho_\textrm{p}} + \frac{1}{\bar{\rho}_\textrm{p}}\right) \right]\delta^\circ_\textrm{p} 
- \frac{1}{12} l^{(a} \bar{m}^{b)} \Theta^+_\textrm{p} \bigg[ 6 \bar{\rho} \Omega^\circ\\
+ 2 \rho (\rho + \bar{\rho}) \frac{\bar{\rho}^2_\textrm{p}}{\rho^4_\textrm{p}} - 3 \frac{(\rho + \bar{\rho})^2}{\rho} \left(\frac{1}{\rho_\textrm{p}} + \frac{1}{\bar{\rho}_\textrm{p}} \right) + 2 \frac{2\rho - \bar{\rho}}{\rho^2} \frac{\bar{\rho}_\textrm{p} (3 \rho_\textrm{p} + \bar{\rho}_\textrm{p})}{\rho_\textrm{p}} + 3 \bar{\rho} (\rho + \bar{\rho}) \left(\frac{1}{\rho^2_\textrm{p}} + \frac{1}{\bar{\rho}^2_\textrm{p}} \right) \bigg] \tau^\circ \delta^\circ_\textrm{p} \\
- \frac{1}{12} l^{(a} \bar{m}^{b)} \Theta^+_\textrm{p} \left[ -2 \rho (\rho + \bar{\rho}) \frac{\bar{\rho}_\textrm{p}}{\rho^4_\textrm{p}} + 3 \bar{\rho} \left(\frac{1}{\rho^2_\textrm{p}} + \frac{1}{\bar{\rho}^2_\textrm{p}} \right) - 2 \frac{2\rho - \bar{\rho}}{\rho^2} \frac{\bar{\rho}_\textrm{p}}{\rho_\textrm{p}} - 3 \bar{\rho} \left(\frac{1}{\rho_\textrm{p}} + \frac{1}{\bar{\rho}_\textrm{p}}\right) \Omega^\circ \right] \tilde{\eth} \delta^\circ_\textrm{p}\\
+ \frac{1}{48} l^a l^b \Theta^+_\textrm{p} \bigg[ 12 \left(\frac{1}{\rho^2_\textrm{p}} + \frac{1}{\bar{\rho}^2_\textrm{p}}\right) - 3 \frac{3 \rho^3 -
12 \rho^2 \bar{\rho} + 5 \rho \bar{\rho}^2 - \bar{\rho}^3}{\rho^2 \bar{\rho}^2} \left(\frac{1}{\rho_\textrm{p}} + \frac{1}{\bar{\rho}_\textrm{p}} \right) - 4 \bar{\rho}^2 \Omega^\circ \frac{\rho_\textrm{p}}{\bar{\rho}^4_\textrm{p}}\\
+ 3 (\rho + \bar{\rho}) \Omega^\circ \frac{4 \rho^4_\textrm{p} - \rho^3_\textrm{p} \bar{\rho}_\textrm{p} - \rho^2_\textrm{p} \bar{\rho}^2_\textrm{p} + \rho_\textrm{p} \bar{\rho}^3_\textrm{p} + \bar{\rho}^4_\textrm{p}}{\rho^2_\textrm{p} \bar{\rho}^3_\textrm{p}(\rho_\textrm{p} + \bar{\rho}_\textrm{p})} - 4 \frac{3 \rho - 4 \bar{\rho}}{\bar{\rho}^2} \Omega^\circ \frac{\rho_\textrm{p}}{\bar{\rho}_\textrm{p}} \bigg] \tilde{\thorn}' \delta^\circ_\textrm{p}\\
+ \frac{1}{24} l^a l^b \Theta^+_\textrm{p} \bigg[ - 6 \frac{\rho^2 - \rho \bar{\rho} + \bar{\rho}^2}{\rho \bar{\rho}} \left(\frac{1}{\rho_\textrm{p}} + \frac{1}{\bar{\rho}_\textrm{p}} \right) - (\rho + \bar{\rho}) \frac{3\rho^2 - 7 \rho \bar{\rho} + 3 \bar{\rho}^2}{\rho^2 \bar{\rho}^2} \left(\frac{\bar{\rho}_\textrm{p}}{\rho_\textrm{p}} + \frac{\rho_\textrm{p}}{\bar{\rho}_\textrm{p}}\right) - 2 \rho^2 \frac{\bar{\rho}_\textrm{p}}{\rho^4_\textrm{p}}\\
- \frac{3\rho^2 - \rho \bar{\rho} + 3 \bar{\rho}^2}{\rho \bar{\rho}} \Omega^{\circ 2} (\rho_\textrm{p} + \bar{\rho}_\textrm{p}) + 6 (\rho + \bar{\rho}) \frac{\rho^4_\textrm{p} - \rho^3_\textrm{p} \bar{\rho}_\textrm{p} + \rho^2_\textrm{p} \bar{\rho}^2_\textrm{p} - \rho_\textrm{p} \bar{\rho}^3_\textrm{p} + \bar{\rho}^4_\textrm{p}}{\rho^3_\textrm{p} \bar{\rho}^3_\textrm{p}} - 2 \bar{\rho}^2 \frac{\rho_\textrm{p}}{\bar{\rho}^4_\textrm{p}} \bigg] \tilde{\eth}' \tilde{\eth} \delta^\circ_\textrm{p}\\
+ \frac{1}{12} l^a l^b \Theta^+_\textrm{p} \bigg[ - 2 \rho^2 \bar{\rho} \bar{\tau}^\circ \frac{\bar{\rho}_\textrm{p}}{\rho^4_\textrm{p}} + 3 \rho \bar{\rho} \bar{\tau}^\circ \left(\frac{1}{\rho^2_\textrm{p}} + \frac{1}{\bar{\rho}^2_\textrm{p}}\right) + \bar{\rho}^2 \bar{\tau}^\circ \frac{\rho^2_\textrm{p}}{\bar{\rho}^4_\textrm{p}} - \rho^2 \bar{\tau}^\circ \frac{\bar{\rho}^2_\textrm{p}}{\rho^4_\textrm{p}} - 4 \rho^2 \bar{\tau}^\circ \frac{\bar{\rho}_\textrm{p}}{\rho^3_\textrm{p}} - 4  \frac{\bar{\rho}}{\rho} \bar{\tau}^\circ \frac{\bar{\rho}_\textrm{p}}{\rho_\textrm{p}}\\
- 3 (\rho + \bar{\rho}) \Omega^\circ \bar{\tau}^\circ \frac{\rho^5_\textrm{p} + 3 \rho^4_\textrm{p} \bar{\rho}_\textrm{p} + \rho_\textrm{p} \bar{\rho}^4_\textrm{p} + \bar{\rho}^5_\textrm{p}}{\rho^2_\textrm{p} \bar{\rho}^2_\textrm{p} (\rho_\textrm{p} + \bar{\rho}_\textrm{p})} + \half (\rho + \bar{\rho}) \frac{3 \rho^2 - 7 \rho \bar{\rho} + 3 \bar{\rho}^2}{\rho^2 \bar{\rho}^2} \bar{\tau}^\circ \frac{\rho^3_\textrm{p} + 3 \rho^2_\textrm{p} \bar{\rho}_\textrm{p} - \rho_\textrm{p} \bar{\rho}^2_\textrm{p} - \bar{\rho}^3_\textrm{p}}{\rho_\textrm{p} \bar{\rho}_\textrm{p}}\\
+ 6 (\rho + \bar{\rho}) \bar{\tau}^\circ \frac{\bar{\rho}^2_\textrm{p}}{\rho^2_\textrm{p} (\rho_\textrm{p} + \bar{\rho}_\textrm{p})} + \half \frac{3 \rho^2 - \rho \bar{\rho} + 3 \bar{\rho}^2}{\rho \bar{\rho}} \Omega^\circ \bar{\tau}^\circ \frac{\rho^3_\textrm{p} + 3 \rho^2_\textrm{p} \bar{\rho}_\textrm{p} + \rho_\textrm{p} \bar{\rho}^2_\textrm{p} + \bar{\rho}^3_\textrm{p}}{\rho_\textrm{p} \bar{\rho}_\textrm{p}} \bigg] \tilde{\eth} \delta^\circ_\textrm{p}\\
+ \frac{1}{12} l^a l^b \Theta^+_\textrm{p} \bigg[ - 2 \rho \bar{\rho}^2 \tau^\circ \frac{\rho_\textrm{p}}{\bar{\rho}^4_\textrm{p}} + 3 \rho \bar{\rho} \tau^\circ \left(\frac{1}{\rho^2_\textrm{p}} + \frac{1}{\bar{\rho}^2_\textrm{p}}\right) - \bar{\rho}^2 \tau^\circ \frac{\rho^2_\textrm{p}}{\bar{\rho}^4_\textrm{p}} + \rho^2 \tau^\circ \frac{\bar{\rho}^2_\textrm{p}}{\rho^4_\textrm{p}} - 4 \bar{\rho}^2 \tau^\circ \frac{\rho_\textrm{p}}{\bar{\rho}^3_\textrm{p}} - 4  \frac{\rho}{\bar{\rho}} \tau^\circ \frac{\rho_\textrm{p}}{\bar{\rho}_\textrm{p}}\\
+ 3 (\rho + \bar{\rho}) \Omega^\circ \tau^\circ \frac{\rho^5_\textrm{p} + \rho^4_\textrm{p} \bar{\rho}_\textrm{p} + 3 \rho_\textrm{p} \bar{\rho}^4_\textrm{p} + \bar{\rho}^5_\textrm{p}}{\rho^2_\textrm{p} \bar{\rho}^2_\textrm{p} (\rho_\textrm{p} + \bar{\rho}_\textrm{p})} - \half (\rho + \bar{\rho}) \frac{3 \rho^2 - 7 \rho \bar{\rho} + 3 \bar{\rho}^2}{\rho^2 \bar{\rho}^2} \tau^\circ \frac{\rho^3_\textrm{p} + \rho^2_\textrm{p} \bar{\rho}_\textrm{p} - 3 \rho_\textrm{p} \bar{\rho}^2_\textrm{p} - \bar{\rho}^3_\textrm{p}}{\rho_\textrm{p} \bar{\rho}_\textrm{p}}\\
+ 6 (\rho + \bar{\rho}) \tau^\circ \frac{\rho^2_\textrm{p}}{\bar{\rho}^2_\textrm{p} (\rho_\textrm{p} + \bar{\rho}_\textrm{p})} - \half \frac{3 \rho^2 - \rho \bar{\rho} + 3 \bar{\rho}^2}{\rho \bar{\rho}} \Omega^\circ \tau^\circ \frac{\rho^3_\textrm{p} + \rho^2_\textrm{p} \bar{\rho}_\textrm{p} + 3 \rho_\textrm{p} \bar{\rho}^2_\textrm{p} + \bar{\rho}^3_\textrm{p}}{\rho_\textrm{p} \bar{\rho}_\textrm{p}} \bigg] \tilde{\eth}' \delta^\circ_\textrm{p}\\
+ \frac{1}{6} l^a l^b \Theta^+_\textrm{p} \Re \bigg[ 6 \rho \bar{\rho} (\rho + \bar{\rho}) \frac{1}{\rho^2_\textrm{p}} + 2 \rho^2 \bar{\rho} \frac{\bar{\rho}^2_\textrm{p}}{\rho^4_\textrm{p}} - 3 (\rho^2 + 4 \rho \bar{\rho} + \bar{\rho}^2) \frac{1}{\rho_\textrm{p}} + 2 \rho^2 \bar{\rho}^2_\textrm{p} \frac{2 \rho_\textrm{p} + \bar{\rho}_\textrm{p}}{\rho^4_\textrm{p}}\\
- 6 (\rho + \bar{\rho}) \rho^2_\textrm{p} \frac{\rho^3_\textrm{p} + 3 \rho^2_\textrm{p} \bar{\rho}_\textrm{p} + 2 \rho_\textrm{p} \bar{\rho}^2_\textrm{p} - 2 \bar{\rho}^3_\textrm{p}}{\bar{\rho}^3_\textrm{p} (\rho_\textrm{p} + \bar{\rho}_\textrm{p})^2} + 4 \frac{\bar{\rho}}{\rho} \bar{\rho}_\textrm{p} \frac{3 \rho_\textrm{p} + \bar{\rho}_\textrm{p}}{\rho_\textrm{p}} - 2 \frac{4 \rho - 3 \bar{\rho}}{\rho^2} \bar{\rho}^2_\textrm{p} \frac{2 \rho_\textrm{p} + \bar{\rho}_\textrm{p}}{\rho_\textrm{p}} \bigg] \tau^\circ \bar{\tau}^\circ \delta^\circ_\textrm{p}\\
+ \frac{1}{24} l^a l^b \Theta^+_\textrm{p} \Re \bigg[ 6 \rho^2 \Psi^\circ \left(\frac{1}{\rho^2_\textrm{p}} + \frac{1}{\bar{\rho}^2_\textrm{p}} \right) - 2 \bar{\rho}^2 \Psi^\circ \frac{\rho^2_\textrm{p}}{\bar{\rho}^4_\textrm{p}} + 2 \rho^2 \Psi^\circ \frac{\bar{\rho}^2_\textrm{p}}{\rho^4_\textrm{p}} - 3 \left(2 \frac{\rho^2}{\bar{\rho}} + 3 \rho - 3 \bar{\rho}\right) \Psi^\circ \left(\frac{1}{\rho_\textrm{p}} + \frac{1}{\bar{\rho}_\textrm{p}}\right)\\
- 48 (\rho + \bar{\rho}) \Omega^\circ \Psi^\circ \frac{2 \rho^5_\textrm{p} + 6 \rho^4_\textrm{p} \bar{\rho}_\textrm{p} + 3 \rho^3_\textrm{p} \bar{\rho}^2_\textrm{p} + \rho^2_\textrm{p} \bar{\rho}^3_\textrm{p} + 6 \rho_\textrm{p} \bar{\rho}^4_\textrm{p} + 2 \bar{\rho}^5_\textrm{p}}{\rho^2_\textrm{p} \bar{\rho}^2_\textrm{p} (\rho_\textrm{p} + \bar{\rho}_\textrm{p})} - 2 \frac{3 \rho - 4 \bar{\rho}}{\bar{\rho}^2} \Psi^\circ \rho_\textrm{p} \frac{\rho_\textrm{p} + 3 \bar{\rho}_\textrm{p}}{\bar{\rho}_\textrm{p}}\\
- 2 \frac{4 \rho - 3 \bar{\rho}}{\rho^2} \Psi^\circ \bar{\rho}_\textrm{p} \frac{3 \rho_\textrm{p} + \bar{\rho}_\textrm{p}}{\rho_\textrm{p}} \bigg] \delta^\circ_\textrm{p}\\
+ \frac{1}{6} l^a l^b \Theta^+_\textrm{p} \Re \bigg[ -\bar{\rho}^2 \Omega^\circ \rho^{\prime \circ} \frac{\rho^2_\textrm{p}}{\bar{\rho}^4_\textrm{p}} + 6 \rho^{\prime \circ} \frac{1}{\rho_\textrm{p}} + 3 (\rho + \bar{\rho}) \Omega^\circ \rho^{\prime \circ} \rho_\textrm{p} \frac{\rho^3_\textrm{p} + 3 \rho^2_\textrm{p} \bar{\rho}_\textrm{p} - 2 \rho_\textrm{p} \bar{\rho}^2_\textrm{p} - 6 \bar{\rho}^3_\textrm{p}}{\bar{\rho}^3_\textrm{p} (\rho_\textrm{p} + \bar{\rho}_\textrm{p})^2}\\
- \frac{3 \rho - 4 \bar{\rho}}{\bar{\rho}^2} \Omega^\circ \rho^{\prime \circ} \rho_\textrm{p} \frac{\rho_\textrm{p} + 3 \bar{\rho}_\textrm{p}}{\bar{\rho}_\textrm{p}} \bigg] \delta^\circ_\textrm{p},
\end{multline}
\begin{multline}
\label{propa:5}
{}^+ X^{ab}_{\bar{m}} := \frac{1}{3} l^{(a} \bar{m}^{b)} \Theta^+_\textrm{p} \left[ \rho (\rho + \bar{\rho}) \frac{2 \rho_\textrm{p} - \bar{\rho}_\textrm{p}}{\rho^4_\textrm{p}} - \frac{2 \rho - \bar{\rho}}{\rho^2} \left(1 + \frac{\bar{\rho}_\textrm{p}}{\rho_\textrm{p}} \right) \right] \delta^\circ_\textrm{p} \\
- \frac{1}{6} l^a l^b \Theta^+_\textrm{p} \left[ \frac{4}{\rho} \left(1 + \frac{\bar{\rho}_\textrm{p}}{\rho_\textrm{p}}\right) - \frac{3 \bar{\rho}}{\rho^2} \left(1 + \frac{\bar{\rho}_\textrm{p}}{\rho_\textrm{p}}\right) + \rho^2 \frac{2 \rho_\textrm{p} - \bar{\rho}_\textrm{p}}{\rho^4_\textrm{p}} - 3 \bar{\rho} \frac{2 \rho^2_\textrm{p} - \bar{\rho}^2_\textrm{p}}{\rho^3_\textrm{p} (\rho_\textrm{p} + \bar{\rho}_\textrm{p})} - 3 \rho  \frac{2 \rho^3_\textrm{p} - \bar{\rho}^3_\textrm{p}}{\rho^4_\textrm{p} (\rho_\textrm{p} + \bar{\rho}_\textrm{p})} \right] \tilde{\eth} \delta^\circ_\textrm{p} \\
- \frac{1}{6} l^a l^b \Theta^+_\textrm{p} \bigg[ - 4 \frac{\bar{\rho}}{\rho} \left(1 + \frac{\bar{\rho}_\textrm{p}}{\rho_\textrm{p}}\right) + \frac{4 \rho - 3 \bar{\rho}}{\rho^2} \frac{\bar{\rho}_\textrm{p}}{\rho_\textrm{p}} (\rho_\textrm{p} + \bar{\rho}_\textrm{p}) + \rho^2 \frac{6 \rho^2_\textrm{p} - 4 \rho_\textrm{p} \bar{\rho}_\textrm{p} - \bar{\rho}^2_\textrm{p}}{\rho^4_\textrm{p}} + 2 \rho^2 \bar{\rho} \frac{2 \rho_\textrm{p} - \bar{\rho}_\textrm{p}}{\rho^4_\textrm{p}}\\
- 3 (\rho + \bar{\rho}) \frac{2 \rho^3_\textrm{p} - 2 \rho_\textrm{p} \bar{\rho}^2_\textrm{p} - \bar{\rho}^3_\textrm{p}}{\rho^3_\textrm{p} (\rho_\textrm{p} + \bar{\rho}_\textrm{p})} \bigg] \bar{\tau}^\circ \delta^\circ_\textrm{p}.
\end{multline}
With these choices, equation \eqref{eq:Xdef} 
gives
$\Theta^+_{\textrm{p}} x_{ab}^{\textrm{hom}}$ after contracting into $T_{a_\textrm{p} b_{\textrm{p}}}$ and integrating over 
$u_\textrm{p},\theta_\textrm{p},\varphi_{_\textrm{p}*}$. Since this gives the retarded solution to the GHZ transport equations for the corrector $x_{ab}$ with the given source $T_{ab}l^b$, 
we therefore obtain the following theorem.
\begin{theorem}
\label{thm:1}
Let $T_{ab}$ be a conserved smooth stress-energy tensor whose support is contained in the causal future of some compact subset (i.e.,
including no horizons or infinities) of Kerr. Then
\begin{equation}\label{eq:corr conv}
x_{ab}^+ = \, {}^+ {\mathcal X}_{ab}{}^{a_{\textrm{p}} b_{\textrm{p}}} * T_{a_{\textrm{p}} b_{\textrm{p}}},
\end{equation}
with the convolution $*$ defined as in \eqref{propa:1}  and with ${}^+ {\mathcal X}_{ab}{}^{a_{\textrm{p}} b_{\textrm{p}}}$ defined as in \eqref{eq:Xdef}, solves the GHZ transport equations \eqref{eq:xmmb1}, \eqref{eq:xnm1} and \eqref{eq:xnn1} sourced by  $T_{ab}$. 
Furthermore, the support of $x_{ab}^+$ is contained in the causal future of the support of $T_{ab}$, so it is a retarded solution.
\end{theorem}

\noindent
{\bf Remark 1.} By construction, the integrations implicit in ${}^+ {\mathcal X}_{ab}{}^{a_{\textrm{p}} b_{\textrm{p}}} * T_{a_{\textrm{p}} b_{\textrm{p}}}$ (equation \eqref{propa:2}) are over a compact domain given that $T_{ab}$ has its support contained in the causal future of some compact set, whereas, for given $y$, ${}^+ {\mathcal X}_{ab}{}^{a_{\textrm{p}} b_{\textrm{p}}}(y,y_\textrm{p})$ has its support in $y_\textrm{p}$ contained in the causal past of $y$, or more precisely along the past null geodesic tangent to $l^a$ emanating from $y$.
These integrations  are therefore  trivially convergent. 

\noindent
{\bf Remark 2.}
By replacing the step $\Theta^+_\textrm{p}$, see equation \eqref{Th+}, everywhere with the opposite step $\Theta^-_\textrm{p}$, defined as
\begin{equation}
\label{Th-}
\Theta^-_\textrm{p}(r) \equiv\left\{
\begin{matrix}
1 & r \le r_\textrm{p}\\
0 & r > r_\textrm{p}
\end{matrix}
\right. \circeq \GHPw{0}{0},
\end{equation}
we obtain an advanced Green’s function ${}^- {\mathcal X}_{ab}{}^{a_{\textrm{p}} b_{\textrm{p}}} $ of the GHZ transport equations.

\noindent
{\bf Remark 3.}
The construction leading to 
${}^\pm {\mathcal X}_{ab}{}^{a_{\textrm{p}} b_{\textrm{p}}}(y,y_\textrm{p})$ gives the Green's function property
\begin{equation}\label{eq:GF eq corr}
\left[ l^b \, (\E {}^\pm {\mathcal X})_{ab}{}^{a_{\textrm{p}} b_{\textrm{p}}} \, l_{b_\textrm{p}}
\right]
(y,y_\textrm{p}) = 
    g_a{}^{a_\textrm{p}} \delta(y,y_{\textrm{p}}),
\end{equation}
which may be seen, in the retarded case, as an equivalent way of expressing theorem \ref{thm:1}. Here, $g_a{}^{a_\textrm{p}}$ is the bi-tensor of parallel transport from $y_{\textrm{p}}$ to $y$.

\noindent
{\bf Remark 4.} The expressions \eqref{eq:Xdef} and \eqref{propa:3}, \eqref{propa:4}, \eqref{propa:5}
entering ${}^\pm {\mathcal X}_{ab}{}^{a_{\textrm{p}} b_{\textrm{p}}}$
simplify in Schwarzschild $(a=0)$, where, in our coordinate system and frame, $\rho=-1/r,\Omega^\circ=\tau^\circ =0$. 
We give the explicit expressions for \eqref{propa:3}--\eqref{propa:5} in Schwarzschild in \ref{app:Schw}.
In Minkowski spacetime, we additionally have $\Psi^\circ = 0$.

\subsection{Frequency domain representation of ${}^\pm {\mathcal X}_{ab}{}^{a_{\textrm{p}} b_{\textrm{p}}}$}
\label{sec:freqdom}

In applications, it is potentially useful to have a frequency domain representation of the Green’s functions 
${}^\pm {\mathcal X}_{ab}{}^{a_{\textrm{p}} b_{\textrm{p}}} $ of the GHZ transport equations. 
Such representations can be obtained from \eqref{propa:3}, \eqref{propa:4} and \eqref{propa:5} 
alternatively in terms of (spin-weighted) spheroidal or spherical harmonics. 
For definiteness, we outline the procedure for spherical harmonics. 

We first substitute into \eqref{propa:3}, \eqref{propa:4} and \eqref{propa:5} the
completeness relation for the $\delta^\circ$ delta-functions \eqref{eq:deltacirc} \footnote{The sum is a shorthand for $\sum_{\ell,m} = \sum_{\ell =|s|}^\infty \sum_{m = -\ell}^\ell$.}, 
\begin{equation}
\begin{split}
\delta^\circ_\textrm{p}(y)
&= \frac{\delta(u-u_\textrm{p})\delta(\theta-\theta_\textrm{p})\delta(\varphi_*-\varphi_{*\textrm{p}})}{
\Sigma_\textrm{p} \sin \theta_\textrm{p}} \\
&= \frac{1}{\Sigma_\textrm{p}} \sum_{\ell,m}\, \int\limits_{-\infty}^\infty \frac{\intd \omega}{2\pi}
e^{-i\omega(u-u_\textrm{p})} Y_{\ell, m}(\theta,\varphi_*) \overline{Y}_{\ell, m}(\theta_\textrm{p},\varphi_{*\textrm{p}}).
\end{split}
\end{equation}
The action of Held's operators $\tilde{\eth}, \tilde{\eth}',\tilde{\thorn}' $ on a mode ${}_sY_{\ell, m}(\theta,\varphi_*)e^{-i\omega u}$ is next inferred from table 
\ref{tab:2}:
\begin{subequations}
\label{eq:Chandop}
\begin{align}
\tilde{\eth} = {}_s \mathcal{L}_{m}^\dagger  -\frac{1}{\sqrt{2}} a\omega \sin \theta, \quad {}_s \mathcal{L}_{m}^\dagger 
 =-\frac{1}{\sqrt{2}} \left( \partial_\theta
- m \csc \theta  - s \cot \theta \right), \\
\tilde{\eth}'   = {}_s \mathcal{L}_{m}  +\frac{1}{\sqrt{2}} a\omega \sin \theta, \quad 
{}_s \mathcal{L}_{m}   = -\frac{1}{\sqrt{2}} \left( \partial_\theta +m \csc \theta + s \cot \theta \right), \\
\tilde{\thorn}'  = -i\omega.
\end{align}
\end{subequations}
Here, the operators ${}_s \mathcal{L}_{m}$ and ${}_s \mathcal{L}_{m}^\dagger$ are spin-lowering respectively spin-raising 
operators for the spin-weighted spherical harmonics ${}_s Y_{\ell, m}$, defined by \cite{Chandrasekhar:1984siy}
\begin{subequations}\label{eq:SpheroidalHarmonicsIdentities}
\begin{align}
{}_{-s} \mathcal{L}^\dagger_{m} \, {}_s Y_{\ell, m} = \sqrt{\frac{(\ell - s)(\ell + s + 1)}{2}} {}_{s+1} Y_{\ell, m},\\
{}_{s} \mathcal{L}_{m} \, {}_s Y_{\ell, m} = -\sqrt{\frac{(\ell + s)(\ell - s + 1)}{2}} {}_{s-1} Y_{\ell, m}
,\label{eq:bar sYlm}
\end{align}
\end{subequations}
and the initial condition ${}_0 Y_{\ell, m} = Y_{\ell, m}$. The Held scalars $\Omega^\circ$ and $\tau^\circ$ appearing in \eqref{propa:3}, \eqref{propa:4} and \eqref{propa:5}
are given by trigonometric functions (see table \ref{tab:2}),  
and $\rho$ \eqref{eq:Heldrho} may trivially be expanded as
\begin{equation}
\label{eq:rhoexp}
\rho = -\frac{1}{r}\left[1+\frac{ia \cos \theta}{r} + O(a^2/r^2)\right]
\end{equation}
since $a/r < 1$.
Products of spin-weighted spherical harmonics and trigonometric functions as appearing 
via \eqref{eq:Chandop} when replacing $\tilde{\eth}, \tilde{\eth}'$ by spin-raising and lowering operators,
or via $\Omega^\circ, \tau^\circ$, or via \eqref{eq:rhoexp} may be converted by applying suitable angular momentum addition rules 
into linear combinations of spin-weighted spherical harmonics. The prerequisite formulas are provided in appendix \ref{app:addthm}.

Carrying out these steps in \eqref{propa:3}, \eqref{propa:4} and \eqref{propa:5} and substituting these into \eqref{eq:Xdef} converts the kernel for ${}^\pm {\mathcal X}_{ab}{}^{a_{\textrm{p}} b_{\textrm{p}}} $ into a mode sum/integral of sums of products of polynomial in $r, r_\textrm{p}$ and spin-weighted 
spherical harmonics of various spin-weights. 

\subsection{Green’s functions for Weyl scalar}
\label{sec:Phiprop1}

The retarded Green’s function giving the retarded Hertz potential $\Phi^+$ as in equation 
\eqref{propa:2} has been constructed by Ori \cite{Ori:2002uv}.  We here recall his method with minor modifications, emphasizing the causal 
character of the Green's function, and providing a more detailed justification of some steps.

According to equation \eqref{eq:th4}, the reconstructed part $\Phi^+$ is obtained  from 
the retarded solution $\psi_0^+$ of the spin $s=+2$ Teukolsky equation, $\O\psi_0^+= {}_{+2}T$. In its turn, $\psi_0^+$
may be obtained from the retarded Green’s function, ${}^+\mathcal{G}(y,y_\textrm{p})$, of the Teukolsky operator $\O$ by the variation of constants method \cite{Leaver}. We now recall this method as a preparation and to set up notation for the next subsection. 

The Teukolsky equation is separable for any spin $s$ in BL coordinates and the Kinnersley frame \cite{Teukolsky:1972my}, 
so we can obtain a solution as a linear superposition of modes,
\begin{equation}
\label{modes}
{}_s \Psi_{\omega \ell m}(t,r,\theta,\varphi) = {}_s R_{\omega \ell m}(r) \, {}_s S_{\ell m}(\theta,\varphi;a\omega) e^{-i\omega t}. 
\end{equation}
Here, ${}_s S_{\ell m}(\theta,\varphi;a\omega)$ are the spin-weighted spheroidal harmonics with harmonic dependence $e^{im\varphi}$.
 Dropping for notational simplicity the mode labels $\omega,\ell,m$ as e.g., in ${}_s R_{\omega \ell m} \equiv {}_s R$, the radial functions obey the Teukolsky equation \cite{Teukolsky:1973ha}
\begin{equation}
\Delta^{-s} (\Delta^{s+1} {}_s R')' - {}_s V {}_s R = -\Sigma \ {}_s T.
\end{equation}
Here we have denoted by ${}_s T_{\omega \ell m}\equiv {}_s T$ 
(in an abuse of notation) the \emph{coefficient} of the source (e.g.,
${}_{+2} T={\mathcal S}^{ab}T_{ab}$ for $s=+2$
) in an expansion in modes \eqref{modes}.
The potential is
\begin{equation}
{}_s V \equiv (iK)^2/\Delta +is K\Delta'/\Delta -2iKs +{}_s \lambda,
\end{equation}
with ${}_s \lambda$ the separation constant, $K \equiv (r^2+a^2)\omega-am$ and $|m| \le \ell \ge |s|$. For gravitational perturbations, it is $s=\pm 2$.  In the following, the mode labels will appear intermittently and are otherwise understood implicitly.

For the retarded Green's function, we require a specific basis of radial mode functions, defined uniquely for $\Im \omega \ge 0, \omega \neq 0$ by the 
conditions, for $s=+2$,
\begin{equation}
\label{eq:R+}
\begin{split}
&{}_{+2} R^{\rm up} \sim \frac{e^{i\omega r_*}}{r^{5}}, \quad r_* \to \infty, \\
&    {}_{+2} R^{\rm in} \sim \frac{e^{-ik r_*}}{\Delta^2}, \quad r_* \to -\infty.
\end{split}
\end{equation}
Here, $k=\omega-m\Omega_+$ is the horizon frequency. 
Similarly, for the advanced Green's function, we require the basis
\begin{equation}
\label{eq:R-}
\begin{split}
&{}_{+2} R^{\rm down} \sim \frac{e^{-i\omega r_*}}{r}, \quad r_* \to \infty, \\
&    {}_{+2} R^{\rm out} \sim e^{ik r_*}, \quad r_* \to -\infty,
\end{split}
\end{equation}
which is defined uniquely for $\Im \omega \le 0,\ \omega \neq 0$.

The response kernel for the $s=+2$ advanced Green's function is defined as \cite{Leaver,Chrzanowski:1975wv,CKO:2016} 
\begin{equation}
\label{eq:g-}
\begin{split}
    & {}^- g_{\omega \ell m}(r,r_\textrm{p}) =  \, \frac{
    \Delta^2_\textrm{p}}{
    {}^- W_{\ell m}(\omega)
    } \times \\
    & \left[
\Theta^+(r-r_\textrm{p})
{}_{+2}R_{\omega \ell m}^\textrm{out}(r_\textrm{p})  {}_{+2}R_{\omega \ell m}^\textrm{down}(r)
    +
\Theta^+(r_\textrm{p}-r)
{}_{+2}R_{\omega \ell m}^\textrm{out}(r)  {}_{+2}R_{\omega \ell m}^\textrm{down}(r_\textrm{p})
    \right]
    ,
\end{split}   
\end{equation}
whereas for the retarded Green's function, the response kernel is
\begin{equation}
\label{eq:g+}
\begin{split}
    &{}^+ g_{\omega \ell m}(r,r_\textrm{p}) =  \, \frac{
    \Delta^2_\textrm{p}}{
    {}^+ W_{\ell m}(\omega)
    }\times \\
    &\left[
\Theta^+(r-r_\textrm{p})
{}_{+2}R_{\omega \ell m}^\textrm{in}(r_\textrm{p})  {}_{+2}R_{\omega \ell m}^\textrm{up}(r)
    +
\Theta^+(r_\textrm{p}-r)
{}_{+2}R_{\omega \ell m}^\textrm{in}(r)  {}_{+2}R_{\omega \ell m}^\textrm{up}(r_\textrm{p})
    \right]
    .
\end{split}   
\end{equation}
Here, $\Theta^+(r)=1$ if $r\ge 0$, $\Theta^+(r)=0$ if $r<0$ (so that, in particular,  $\Theta^+(r-r_\textrm{p})=\Theta^+_\textrm{p}(r)$), and
the Wronskian is given by 
\begin{equation}
    {}^- W = \Delta^3 \left( 
    {}_{+2} R^{\rm down} \frac{\intd}{\intd r} {}_{+2} R^{\rm out} -
    {}_{+2} R^{\rm out} \frac{\intd}{\intd r} {}_{+2} R^{\rm down}
    \right) 
\end{equation}
for the (down, out) in modes, $(-)$, and by a similar expression $(+)$ for the (up, in)  modes.
These Wronskians cannot vanish for real $\omega$ (and for $\Im \omega < 0$ respectively $\Im \omega > 0$) due to the absence of 
anti-quasi-normal mode respectively quasi-normal mode frequencies on (and below respectively above) the real axis \cite{Andersson,Whiting}. The response kernels \eqref{eq:g+}, \eqref{eq:g-} satisfy
\begin{equation}
    \left[\Delta^{-2} \partial_r (\Delta^3 \partial_r) - {}_{+2} V\right] {}^\pm g(r,r_{\textrm{p}}) = -\delta (r-r_{\textrm{p}}).
\end{equation}

The spacetime retarded $(+)$ and advanced $(-)$ Green's functions are by definition constructed so that
\begin{equation}
\label{psi0}
\psi^\pm_0(y) = {}^\pm\mathcal{G} *  {}_{+2}T(y)
\end{equation}
are retarded and advanced solutions to the sourced (by ${}_{+2}T={\mathcal S}^{ab}T_{ab}$) spin $s=+2$ Teukolsky equations, 
where the convolution notation $*$ is analogous to equation \eqref{propa:1}. 
Due to the stationary character of the Kerr metric, these Green's functions 
only depend on the times via $t-t_\textrm{p}$. They 
are uniquely defined by 
\begin{equation}\label{eq:GF eq psi0}
\O \, {}^\pm\mathcal{G}(y,y_{\rm p}) = \delta(y,y_{\rm p})
\end{equation}
once we require that their time-domain support should be 
\begin{equation}
\label{eq:Gsupp}
{}^\pm\mathcal{G}(t,r,\theta,\varphi;t_\textrm{p},r_\textrm{p}, \theta_\textrm{p},\varphi_\textrm{p}) = 0 
\quad 
\text{when}
\begin{cases}
t<t_\textrm{p} & \text{for} \,  (+), \\
t>t_\textrm{p} & \text{for} \, (-).
\end{cases}
\end{equation}
Given that solutions of Teukolsky's equation with compactly supported sources decay in time \cite{DHR}, we may 
take a Fourier-Laplace transform of the retarded and advanced Green's functions in the relative time $t-t_\textrm{p}$ at liberty. 
In fact, as a consequence of \eqref{eq:Gsupp}, this Fourier-Laplace transform must 
be analytic for $\Im \omega >0$ and go to zero as $\Im \omega \to \infty$ for the retarded Green's function, 
whereas it must be analytic for $\Im \omega <0$ and go to zero as $\Im \omega \to -\infty$ for the advanced Green's function. 
This formally leads to \cite{Leaver,CKO:2016}
\begin{equation}
\label{eq:Gretadv}
{}^\pm\mathcal{G}(y,y_\textrm{p}) = \sum_{\ell,m}\; \int\limits_{-\infty\pm i0}^{\infty\pm i0} \intd \omega \, e^{-i\omega(t-t_{\textrm{p}})} \
{}^\pm g_{\omega \ell m}(r,r_\textrm{p})
{}_{+2}S_{\ell m}(\theta, \varphi,a\omega)   {}_{+2}\overline{S}_{\ell m}(\theta_\textrm{p}, \varphi_{\textrm{p}},a\omega).
\end{equation}
The forms \eqref{eq:g+} and \eqref{eq:g-} of the response kernels ${}^\pm g_{\omega \ell m}$  follow from the analyticity and decay properties of the Fourier-Laplace transform
which enforce the required oscillatory behaviors of the mode functions, i.e. up and in  modes for the retarded $(+)$ 
respectively down and out modes for the advanced $(-)$ response kernels; see \eqref{eq:R+} respectively \eqref{eq:R-}. 

A  rigorous argument for \eqref{eq:Gretadv}, \eqref{eq:g+} and \eqref{eq:g-} -- assuming global decay in time of the solutions to the Teukolsky equation as later proven in \cite{DHR} -- was given for $s=0$ and Schwarzschild by 
\cite{FredenhagenHaag}. This case is special in that: (i) the angular modes are spherical- instead of spin-weighted spheroidal harmonics, which, contrary to the former, depend on $\omega$; (ii) The potential $_s{}V$ is real and sign-definite for real $\omega$ when $s=0$ in Schwarzschild, but not for $s \neq 0$ or in Kerr; (iii) The potential is short range for $s=0$, i.e., it decays at least as fast as $O(1/r_*^2)$ (it actually decays exponentially) for $r_* \to -\infty$ and as $O(1/r^{2})$ for $r \to +\infty$, but it is not short range for $s \neq 0$. Nevertheless, the 
arguments by \cite{FredenhagenHaag} can be generalized to Kerr and $s\neq 0$: (i) The separation constant ${}_s \lambda$ and spheroidal harmonics $_s S_{\ell m}$ have an analytic dependence on $\omega$ except for cuts starting at at most finitely many $\omega$ values (see e.g., \cite[Prop. 2.1]{TdC20} and references therein); (ii) Even though the potential fails to be sign definite for real $\omega$, one may still establish the absence of anti-quasi-normal mode respectively quasi-normal mode frequencies on (and below respectively above) the real axis \cite{Andersson,Whiting}, and this is what matters to the argument; (iii) The radial equation may be cast into a standard Schr\" odinger-type equation in terms of a  
new variable\footnote{In fact, this variable is given by 
$\intd \eta = (\gamma(r^2+a^2)/\Delta) \intd r$, where $\gamma$ is the function appearing in the Sasaki-Nakamura transformation theory; see footnote **) on p.~1794 of \cite{Sasaki1}. Note that $\gamma = O(1)$ as $r_* \to \pm \infty$; see 
\cite[App. A]{Sasaki1}.} $\eta$ and a new potential ${}_s U(\eta)$ with the help of a Sasaki-Nakamura transformation \cite{Sasaki,Sasaki1}. After such a transformation, 
the potential decays exponentially for $\eta \to -\infty$ and as 
${\rm const}/\eta^2$ for $\eta \to +\infty$. The asserted representations \eqref{eq:Gretadv} then follow by the same arguments as in \cite{FredenhagenHaag}. 

\subsection{Green’s functions for Hertz potential}
\label{sec:Phiprop}

The Weyl scalar $\psi_0^\pm$ is related to the Hertz potential $\Phi^\pm$ through $-4\psi_0^\pm = \th^4 \bar \Phi^\pm$, to which we should find solutions $\Phi^\pm$ subject to retarded respectively advanced boundary conditions, i.e. the support of $\Phi^\pm$ should be contained in the causal future respectively past of that of $\psi_0^\pm$. Since $\thorn$ is basically a directional derivative tangent to the orbits of the principal null direction $l^a$ ($=(\partial_r)$ in retarded KN coordinates and the Kinnersley frame), it follows that $-4\psi_0^\pm = \th^4 \bar \Phi^\pm$ is an ODE along these null geodesics. This ODE must, therefore, be solved for $\Phi^\pm$ with retarded respectively advanced boundary conditions
with respect to the affine time-parameter along the geodesic. Since 
$\psi_0^\pm$ itself is given by equation \eqref{psi0}, we will thereby obtain a formula for $\Phi^\pm$ in terms of a new retarded $(+)$ or advanced $(-)$ Green's function, $^\pm \mathcal{H}$. It is related \footnote{It can also be shown, by using Eq.~\eqref{eq:SE=OT}, that it is a Green function of the self-adjoint operator $\mathcal{S}\mathcal{E}\mathcal{S}^\dagger$, i.e.,
$\mathcal{S}\mathcal{E}\mathcal{S}^\dagger \, {}^\pm \mathcal{H}(y,y_{\rm p}) = \delta(y,y_{\rm p})$.} to the advanced and retarded Green's functions ${}^\pm\mathcal{G}$ of Teukolsky's equation for $s=+2$ by
\begin{equation}
\label{eq:GP4H}
    -\tfrac{1}{4} \thorn^4 \, {}^\pm \mathcal{H}(y,y_\textrm{p}) = 
    {}^\pm\mathcal{G}(y,y_\textrm{p})
\end{equation}
and defined uniquely by the property that the $y$-support of ${}^\pm \mathcal{H}(y,y_\textrm{p})$
be contained in the causal past $(+)$ respectively future $(-)$ of 
$y_{\textrm{p}}$.

However, following \cite{Ori:2002uv}, instead of finding ${}^\pm \mathcal{H}(y,y_\textrm{p})$ by directly integrating \eqref{eq:GP4H} along the orbits of $l^a$ subject to these 
boundary conditions, it appears easier to rely on a suitable application of the Teukolsky-Starobinsky identities (see e.g., \cite{Chandrasekhar:1984siy}). 
These imply that if ${}_{-2}R_{\omega \ell m}(r)$ is a solution to the 
radial Teukolsky equation (in BL coordinates and the Kinnersley frame) for spin $s=-2$, then 
\begin{equation}
\label{eq:D4}
    {}_{+2}R_{\omega \ell m}(r) =
    \left( \prod_{s=-2}^{1} {}_{s} {\mathcal D}_{\omega m} \right)
    {}_{-2}R_{\omega \ell m}(r) 
\end{equation}
is a solution for spin $s=+2$. Here, ${}_{s} {\mathcal D}_{\omega m}$ is the radial differential operator that implements the action of $\th$ on a mode of spin $s$ in this frame and coordinate system,
\begin{equation}
\label{eq:ThtoD}
    \begin{split}
        \thorn \Big[ {}_s R(r) \, {}_s S(\theta,\varphi) \, e^{-i\omega t}\Big] =& \Big[ {}_s {\mathcal D}_{} \, {}_s R(r) \Big]
        {}_s S(\theta,\varphi) \, e^{-i\omega t},\\
        \thorn' \Big[ {}_s R(r) \, {}_s S(\theta,\varphi) \, e^{-i\omega t}\Big] =& \Big[ {}_s {\mathcal D}_{}^\dagger \, {}_s R(r) \Big]
        {}_s S(\theta,\varphi) \, e^{-i\omega t},
    \end{split}
\end{equation}
where we have omitted the dependence on the mode indices $m,\ell,\omega$
on ${}_{s} {\mathcal D}_{\omega m}$ and in ${}_{s}S_{\ell m}(\theta, \varphi,a\omega)$ for notational simplicity.
Explicitly, we have
\begin{equation}
    {}_s {\mathcal D}_{\omega m} = \partial_r -iK/\Delta, \quad 
    {}_s {\mathcal D}_{\omega m}^\dagger = \partial_r +iK/\Delta,
\end{equation}
so it happens to be actually independent of $s$ in our given frame, and we will consequently write it as 
${\mathcal D}_{\omega m}$ in the following.

Consider the canonically normalized up, down, in, out modes for spin $s=-2$ defined by the asymptotic conditions 
\begin{equation}
\label{eq:538}
    {}_{-2} R^{\rm up} \sim \frac{e^{i\omega r_*}}{r^{3}} \quad (\Im \omega >0)
    , \quad 
    {}_{-2} R^{\rm down} \sim \frac{e^{-i\omega r_*}}{r} \quad (\Im \omega <0)
\end{equation} 
for $r_* \to \infty$, as well as 
\begin{equation}
\label{eq:539}
    {}_{-2} R^{\rm in} \sim \Delta^2 e^{-ik r_*} \quad  (\Im \omega >0), \quad
    {}_{-2} R^{\rm out}(r) \sim e^{ik r_*} \quad (\Im \omega <0)
\end{equation} 
as $r_* \to -\infty$, where $k=\omega-m\Omega_+$. Ref.~\cite{Ori:2002uv} has determined the constants $C^{(i)}_{\omega \ell m}$, $i \in \{ \text{in, out, up, down} \}$, such that
  \begin{equation}
    {}_{+2}R_{\omega \ell m}^{(i)} =
 C^{(i)}_{\omega \ell m} {\mathcal D}_{\omega m}^4
    {}_{-2}R_{\omega \ell m}^{(i)}.
    \label{TeukStar}
\end{equation}
Their values are \cite{Ori:2002uv}
\begin{subequations}
\label{lem:Const}
\begin{align}
C^{{\rm in}}_{\omega \ell m} =& \frac{1}{\prod_{s=2}^{-1}[4(\omega-m\Omega_+)Mr_+ + is(r_+-r_-)]},\\
C^{{\rm out}}_{\omega \ell m} =& \frac{\prod_{s=2}^{-1}[4(\omega-m\Omega_+)Mr_+ - is(r_+-r_-)]}{16(D_{\ell m}(a\omega)^2 + 9 M^2 \omega^2)},\\
C^{{\rm up}}_{\omega \ell m} =& \frac{\omega^4}{D_{\ell m}(a\omega)^2 + 9 M^2 \omega^2},\\
C^{{\rm down}}_{\omega \ell m} =& \frac{1}{16\omega^4}.
\end{align}
\end{subequations}
Here, $D_{\ell m}(a\omega)$ is the spin $s=2$ angular Teukolsky-Starobinsky constant whose value may be found, e.g., in 
\cite{Chandrasekhar:1984siy}.

We note that the Teukolsky-Starobinsky identities are generally valid for $\omega$ complex. At the so-called algebraically-special complex frequencies, though, these identities break down~\cite{Chandrasekhar84}.
Furthermore, $D_{\ell m}(a\omega)$ depends on the separation constant ${}_s\lambda$, which, while it has branch points in the complex $\omega$-plane, these lie {\it off} the real axis (see, e.g.,~\cite{hunter1982eigenvalues} for a proof in the scalar $s=0$ case, which is readily extendible to $s\neq 0$ as well). Since the $\omega$-integral in Eq.~\eqref{eq:Gretadv} can be taken {\it arbitrarily} close to the real axis, both the algebraically-special frequencies and the branch points can be avoided. Similarly, the up and down solutions are not well-defined for $\omega=0$ and neither are the in and out solutions for $k=0$; see, e.g.,~\cite{van2015metric} for the Teukolsky-Starobinsky identities at these  frequencies. However,  the $\omega$-integral in Eq.~\eqref{eq:Gretadv} is {\it off} the real axis, thus avoiding these particular frequencies.

Given that 
${\mathcal D}_{\omega m}$ is the radial differential operator that implements the action of $\th$ on a mode of any spin $s$, 
one is looking for a response function ${}^\pm h_{\omega \ell m}(r,r_\textrm{p})$ such that 
\begin{equation}
\label{eq:DGP}
    -\tfrac{1}{4}  {\mathcal D}_{\omega m}^4
    {}^\pm h_{\omega \ell m}(r,r_\textrm{p}) = {}^\pm g_{\omega \ell m}(r,r_\textrm{p}),
\end{equation}
with the idea that the corresponding spacetime kernel
\begin{equation}
\label{eq:Hdef}
{}^\pm\mathcal{H}(y,y_\textrm{p}) = \sum_{\ell,m} \int\limits_{-\infty\pm i0}^{\infty\pm i0} \intd \omega\, e^{-i\omega(t-t_{\textrm{p}})} \
{}^\pm h_{\omega \ell m}(r,r_\textrm{p})
{}_{+2}S_{\ell m}(\theta, \varphi,a\omega)   {}_{+2} \overline{S}_{\ell m}(\theta_\textrm{p}, \varphi_{\textrm{p}},a\omega)
\end{equation}
gives the retarded respectively advanced Hertz potentials $\Phi^\pm$
satisfying $-4\psi_0^\pm = \th^4 \bar \Phi^\pm$ and $\O \psi_0 = {}_{+2}T$, i.e. 
such that equation \eqref{eq:GP4H} holds with the prerequisite retarded and advanced boundary conditions on ${}^\pm\mathcal{H}(y,y_\textrm{p})$ described below equation~\eqref{eq:GP4H}.

Equation \eqref{eq:DGP} may be solved using \eqref{TeukStar} and \eqref{lem:Const}. To pick out the solution giving rise to the correct causal support of $^\pm \mathcal{H}$, we must know the general solution to the homogeneous equation involving the fourth-order differential operator in equation \eqref{eq:DGP}. It turns out to be given by \cite{Ori:2002uv}
    \begin{equation}
    \label{eq:homsoln}
        {}_{-2} R^\textrm{hom}_{\omega \ell m}(r) = \exp \left( i\omega r_*  -im\frac{a}{r_+-r_-} \log\frac{r-r_+}{r-r_-} \right) \sum_{j=0}^3 B_j (r - r_\textrm{p})^j, 
    \end{equation}
    where  $B_j, j= 0, 1, 2, 3$ are free constants. These homogeneous solutions  behave as 
   \begin{equation}
    \label{eq:homsoln1}
        {}_{-2} R^\textrm{hom} \sim 
        {\rm const} 
        \begin{cases}
            (\sum_{j=0}^3 B_j r^j) e^{i\omega r_*} & \text{$r_* \to \infty$}, \\
            e^{ik r_*} & \text{$r_* \to -\infty$},
        \end{cases} 
    \end{equation}
so they oscillate in the same way as the spin $s=-2$ up respectively out modes at infinity respectively the horizon. 
Since the causal time domain properties of ${}^\pm \mathcal{H}$ imply that  the corresponding response kernel
 \eqref{eq:D4} must not grow exponentially for $\Im \omega \to \pm \infty$, and since the form of the response kernel 
 is fixed up to adding homogeneous solutions, the only freedom 
 is to add solutions to the homogeneous equations 
 in a specific way which does not destroy these properties of ${}^\pm h_{\omega \ell m}(r,r_\textrm{p})$. Furthermore, ${}^\pm B_{j \omega \ell m}(r_\textrm{p}), j= 0, 1, 2, 3$ must be chosen as functions of $r_\textrm{p}$ that are determined uniquely by the requirement that ${}^\pm h_{\omega \ell m}(r,r_\textrm{p})$ are four times continuously differentiable at $r=r_\textrm{p}$, which is necessary and sufficient to ensure the validity of equation \eqref{eq:DGP} at $r=r_\textrm{p}$.

The final forms of the response kernels that  uniquely follow from these requirements
are as follows, where for simplicity we omit the mode labels $\omega,\ell,m$:
\begin{equation}
\label{eq:h+1}
\begin{split}
& {}^- h(r,r_\textrm{p}) =  -\, \frac{4\Delta_{\textrm{p}}^2}{{}^- W} \times \\
\bigg[
& C^\textrm{down} \Theta^+(r-r_\textrm{p}) {}_{-2}R^\textrm{down}(r) {}_{+2}R^\textrm{out}(r_\textrm{p}) + C^\textrm{out}  \Theta^+(r_\textrm{p}-r) {}_{-2}R^\textrm{out}(r) {}_{+2}R^\textrm{down}(r_\textrm{p})
\bigg] + \\
& -\, \frac{4\Delta_{\textrm{p}}^2}{{}^- W} \exp\left\{
i\omega(r_*-r_{*\textrm{p}}) - \frac{ima}{r_+-r_-}
\log\left[ \frac{(r-r_+)(r_{\textrm{p}}-r_-)}{(r-r_-)(r_{\textrm{p}}-r_+)} \right]
\right\} \times \\
& \Theta^+(r_\textrm{p}-r) \sum_{j=0}^3 \frac{(r-r_{\textrm{p}})^j}{j!} \bigg[
C^\textrm{down} {}_{+2}R^\textrm{out}(r_\textrm{p}) \mathcal{D}^j {}_{-2}R^\textrm{down}(r_\textrm{p}) - 
C^\textrm{out}  {}_{+2}R^\textrm{down}(r_\textrm{p}) \mathcal{D}^j {}_{-2}R^\textrm{out}(r_\textrm{p})
\bigg]
\end{split}   
\end{equation}
respectively
\begin{equation}
\label{eq:h-1}
\begin{split}
& {}^+ h(r,r_\textrm{p}) =  -\, \frac{4\Delta_{\textrm{p}}^2}{{}^+ W} \times \\
\bigg[
& C^\textrm{up} \Theta^+(r-r_\textrm{p}) {}_{-2}R^\textrm{up}(r) {}_{+2}R^\textrm{in}(r_\textrm{p}) + C^\textrm{in}  \Theta^+(r_\textrm{p}-r) {}_{-2}R^\textrm{in}(r) {}_{+2}R^\textrm{up}(r_\textrm{p})
\bigg] + \\
& -\, \frac{4\Delta_{\textrm{p}}^2}{{}^+ W} \exp\left\{
i\omega(r_*-r_{*\textrm{p}}) - \frac{ima}{r_+-r_-}
\log\left[ \frac{(r-r_+)(r_{\textrm{p}}-r_-)}{(r-r_-)(r_{\textrm{p}}-r_+)} \right]
\right\} \times \\
& \Theta^+(r-r_\textrm{p}) \sum_{j=0}^3 \frac{(r-r_{\textrm{p}})^j}{j!} \bigg[
C^\textrm{in} {}_{+2}R^\textrm{up}(r_\textrm{p}) \mathcal{D}^j {}_{-2}R^\textrm{in}(r_\textrm{p}) - 
C^\textrm{up}  {}_{+2}R^\textrm{in}(r_\textrm{p}) \mathcal{D}^j {}_{-2}R^\textrm{up}(r_\textrm{p})
\bigg].
\end{split}   
\end{equation}

\subsection{Radiation gauge Green’s functions for metric perturbation}
\label{sec:Mprop}

With the advanced and retarded Green's functions for the Hertz potential ${}^\pm \mathcal{H}(y,y_\textrm{p})$ at hand, and with ${}^\pm \mathcal{X}_{ab}{}^{a_\textrm{p} b_\textrm{p}}(y,y_\textrm{p})$, the advanced and retarded Green's functions for the corrector, having already been obtained in the previous subsection, 
we can now simply put them together in order to get the corresponding Green's functions in IRG for the linearized Einstein operator, $\E$:

\begin{theorem}
   \label{thm:prop}
   The formulas 
   \begin{equation}
   \label{eq:finalprop}
       {}^\pm \mathcal{G}_{ab}{}^{a_\textrm{p} b_\textrm{p}}(y,y_\textrm{p}):=
       {\rm Re} \bigg[ {\mathcal S}^\dagger_{ab} \, {}^\pm \overline{\mathcal{H}}(y,y_\textrm{p}) \, 
       \overline{\mathcal{S}}^{a_\textrm{p} b_\textrm{p}} \bigg] +
       {}^\pm \mathcal{X}_{ab}{}^{a_\textrm{p} b_\textrm{p}}(y,y_\textrm{p})
   \end{equation}
define retarded $(+)$ and advanced $(-)$ Green's functions for the 
linearized Einstein operator, where $\S^\dagger_{ab}$ is the metric reconstruction operator \eqref{eq:Sdag}, where ${}^\pm \mathcal{H}(y,y_\textrm{p})$
is defined through equations \eqref{eq:Hdef}, \eqref{eq:h-1} and \eqref{eq:h+1}, and where ${}^\pm \mathcal{X}_{ab}{}^{a_\textrm{p} b_\textrm{p}}(y,y_\textrm{p})$ is defined through equations \eqref{eq:Xdef}, \eqref{propa:3}, \eqref{propa:4} and \eqref{propa:5}.

This means that, for any conserved stress tensor $T_{ab}$ whose support 
is contained in the causal future $(+)$ respectively past $(-)$ of some compact set, the fields
\begin{equation}
\label{eq:convolution}
    h^\pm_{ab} = \, {}^\pm \mathcal{G}_{ab}{}^{a_\textrm{p} b_\textrm{p}}
    * T_{a_\textrm{p} b_\textrm{p}},
\end{equation}
with $*$ meaning convolution as in equation \eqref{propa:1}, satisfy the following conditions:
\begin{itemize}
    \item they satisfy the linearized EEs, i.e. $(\E h^\pm)_{ab} = T_{ab}$,
    \item they are in IRG, i.e. $h^\pm_{ab}l^b=0$,
    \item their support is contained in the causal future $(+)$ respectively past $(-)$ of the same compact subset.
\end{itemize}
\end{theorem}

\noindent
{\bf Remark 1.} Corresponding forms of the Green's functions for the ORG may be obtained by applying a GHP priming operation to all quantities.
Also note that a different way of stating the last property is that 
${}^\pm \mathcal{G}_{ab}{}^{a_\textrm{p} b_\textrm{p}}(y,y_\textrm{p})$ can be non-zero at most if there is a future- ($+$) or past ($-$) directed 
timelike or null curve from $y_{\rm p}$ to $y$. This follows directly from the constructions in the previous sections. 

\noindent
{\bf Remark 2.} As written, ${}^\pm \mathcal{G}_{ab}{}^{a_\textrm{p} b_\textrm{p}}(y,y_\textrm{p})$ is a differential operator: due to the presence of $\overline{\mathcal{S}}^{a_\textrm{p} b_\textrm{p}}$ in equation~\eqref{eq:finalprop}, ${}^\pm\mathcal{G}_{ab}{}^{a_\textrm{p} b_\textrm{p}}$ acts on smooth rank-two tensor fields not only through convolution but through differentiation. However, we can re-express it in a more conventional form when it acts on test fields with compact support or sufficiently rapid decay properties. This is straightforwardly done through integration by parts:
\begin{equation}
    \int  {\rm Re} \big[ {\mathcal S}^\dagger_{ab} \, {}^\pm \overline{\mathcal{H}}(y,y_\textrm{p}) \, 
       \overline{\mathcal{S}}^{a_\textrm{p} b_\textrm{p}}  T_{a_{\textrm{p}} b_{\textrm{p}}}(y_{\textrm{p}})\big] \, \intd V_{\textrm{p}} =     \int  {\rm Re} \big[{\mathcal S}^\dagger_{ab}\overline{\mathcal{S}}^\dagger_{a_\textrm{p} b_\textrm{p}} {}^\pm \overline{\mathcal{H}}(y,y_\textrm{p})T^{a_{\textrm{p}} b_{\textrm{p}}}(y_{\textrm{p}})\big] \, \intd V_{\textrm{p}}
\end{equation}
for real-valued test fields $T_{a_\textrm{p} b_\textrm{p}}$. (We have swapped the location of contravariant and covariant indices for visual clarity and to emphasise the symmetrical structure of the Green's function below.) Therefore, as an operator on test fields,
   \begin{equation}\label{eq:GF MP}
       {}^\pm \mathcal{G}_{ab a_\textrm{p} b_\textrm{p}}(y,y_\textrm{p})=
       {\rm Re} \bigg[ {\mathcal S}^\dagger_{ab}\overline{\mathcal{S}}^{\dagger}_{a_\textrm{p} b_\textrm{p}} {}^\pm \overline{\mathcal{H}}(y,y_\textrm{p}) 
        \bigg] +
       {}^\pm \mathcal{X}_{ab a_\textrm{p} b_\textrm{p}}(y,y_\textrm{p}).
   \end{equation}
This has the more familiar form of a Green's function that only requires convolution with a source function to yield a solution.

\noindent
{\bf Remark 3.} From an analytical perspective, the retarded solutions in IRG do not have favorable decay properties near $\sI^+$: As one may see e.g., from the integration scheme provided by \cite{Green:2019nam}, in such a gauge, the 
retarded Hertz potential $\Phi^+$ is of order $O(r^3)$ in retarded KN coordinates and the Kinnersley frame, whereas the decay of the retarded corrector $x^+_{ab}$ 
is $x_{nn}^+ = O(r^3), x_{nm}^+ = O(r^3), x_{m\bar m}^+ =O(r)$. The retarded metric perturbation inherits 
correspondingly bad (i.e. non-Bondi type) behaviors since $h^+_{ab} = \Re (\S^\dagger \Phi^+)_{ab}+x_{ab}^+$. To remedy these issues, one has to change to a better-behaved 
gauge in practice, such as the no-string \cite{HT1} or Lorenz \cite{Dolan,Dolan2,Green2} gauges. 

\noindent
{\bf Remark 4.} 
It is \emph{not} possible for our IRG Green's functions to satisfy the Green's identity $(\E^{\rm IRG}\, {}^\pm \mathcal{G})_{ab}{}^{a_\textrm{p} b_\textrm{p}}(y,y_\textrm{p})
 = g_a{}^{a_\textrm{p}} g_b{}^{b_\textrm{p}}\delta(y,y_\textrm{p})$ where $\E^{\rm IRG}$ is the linearized Einstein operator in IRG, and $g_a{}^{a_\textrm{p}}$ is the bi-tensor of parallel transport from $y_\textrm{p}$
 to $y$. To see this, note that $\E^{\rm IRG}$ applied to any symmetric tensor automatically gives a \emph{conserved} tensor, as $(\E^{\rm IRG} h)_{ab} = (\E h^{\rm IRG})_{ab}$, where $h^{\rm IRG}_{ab}$ is the tensor obtained from $h_{ab}$ by setting to zero any $l$ NP component, and as the image of $\E$ is in the subspace of conserved tensors. The theorem \emph{only} shows that the Green's identity holds when both sides are applied to a \emph{conserved} symmetric covariant rank-two tensor, and for this reason, we still call $^\pm \mathcal{G}_{ab}{}^{a_\textrm{p} b_\textrm{p}}$ Green's functions.
Note that the situation is different in Lorenz gauge where the corresponding Green's function satisfies Green's identity even for non-conserved sources. Of course, what we care about physically are conserved sources, so this remark imposes no restrictions on the applicability of our Green's functions in physically relevant applications.

\medskip

Even though conserved stress tensors supported in the causal past respectively future of some compact set are technically convenient and 
may be constructed by a cutoff procedure in Kerr \cite{HT2}
(starting, e.g., with the stress tensor of a point-particle on an eternal bound orbit in Kerr), 
they can have unphysical properties -- propagating negative energy densities -- originating from the initial ``turning on'' respectively final ``turning off'' regions. To avoid this artificial 
situation one can consider a generalization of the setting of theorem \ref{thm:prop}: For the retarded solutions $h_{ab}^+$, we consider ``zero'' initial conditions on $\sI^-$ and 
$\sH^-$,  
corresponding
to the absence of incoming radiation through  $\sI^-$, $\sH^-$ and $i^-$. At $\sI^-$, these are defined by the absence of the leading terms in the usual Bondi expansion, and they
correspond to one order faster decay in $1/r$ than \eqref{eq:hdec} in retarded KN coordinates after applying a GHP-priming operator to all quantities. At $\sH^-$, we simply require $h_{ab}=0$. %

It is plausible that a solution of this characteristic initial value problem will exist if the stress tensor satisfies 
decay rates towards $\sI^-$, $\sH^-$ and $i^-$ corresponding to the absence of 
incoming matter fluxes from the \emph{infinite} past. 
Thus, we expect that we need one 
order faster decay than \eqref{Tdec} at $\sI^-$ (in retarded KN coordinates after applying a GHP-priming operator to all quantities), 
while at $\sH^-$, one should require \eqref{Tdec1} (in advanced KN coordinates). Thus, if 
\begin{equation}
T_{ll},T_{ln}, T_{m\bar m}= O\left(\frac{1}{r^3}\right), \quad
T_{lm}, T_{nm}, T_{mm} = O\left(\frac{1}{r^4}\right), 
\quad T_{nn} = O\left(\frac{1}{r^5}\right)
  \quad \text{at $\sI^-$,}
\end{equation}
if
\begin{equation}
 T_{nn},T_{nm},T_{nl}= O(r-r_+), \quad 
T_{lm}, T_{mm}, T_{ll}, T_{m\bar m} = O(1) \quad
\text{at $\sH^-$,}
\end{equation}
in a NP frame that is well behaved at both $\sI^-$ and $\sH^-$, and if we have ``sufficient''\footnote{We skip a detailed discussion of the necessary decay conditions at $i^-,i^0$, and the bifurcation surface.} 
decay of $T_{ab}$ towards $i^-$, 
then a solution $h_{ab}^+$ to the characteristic initial value problem with trivial data at 
$\sI^-$, $\sH^-$ and $i^-$ ought to exist, and furthermore it will be given by the formula \eqref{eq:convolution}. 
For a matter stress-energy tensor not decaying towards $i^-$, such as \eqref{DetweilerSET} below, producing stationary pieces in the perturbation $h^+_{ab}$, these zero modes (see appendix \ref{app:D}) will need to be added onto our Green's function formula in a manner described in \cite{HT1}: the perturbed mass and spin $\delta M, \delta a$ of the zero mode \eqref{zeromode1}, \eqref{zeromode2} is determined from $T_{ab}$ by Abbott-Deser integrals \cite{Abbott:1981ff}.
Similar remarks should apply to advanced solutions. 

Finally, in Table \ref{table:summary}, we summarize the  equations and expressions for the various fields and corresponding  Green's functions which we deal with in this paper.

\begin{adjustbox}{angle=90, label={table:summary},caption={Summary for various field quantities of interest of the corresponding: field equation (Eqs.~\eqref{eq:Teuk psi0}, \eqref{eq:th4}, \eqref{eq:corr eq} and \eqref{eq:Eh=T}),  Green's function (GF) equation (Eqs.~\eqref{eq:GF eq psi0}, \eqref{eq:GP4H} and \eqref{eq:DGP}, \eqref{eq:GF eq corr} and Remark 4 within this subsection),  expression for the field (in terms of a convolution of the GF with the source; Eqs.~\eqref{psi0}, \eqref{propa:2}, \eqref{propa:1} and \eqref{eq:convolution}) and equation number containing an expression for the GF. For compactness and clarity, here we omit the
%tensorial indices and the 
spacetime points $y$ and $y_\textrm{p}$ in the 
equations (we refer the reader to the equations in the main text for the expressions including the spacetime points).
%full, detailed version of the equations).
The symbol 
$\stackrel{!}{=}$
means that the equality  only holds when applied to a conserved symmetric covariant rank-two tensor.}, float=table}
%\rotatebox{90}{
%\Rotatebox{90}{
%\begin{landscape}
%\begin{table}[h!]
%\begin{sidewaystable}
%\ctable[
%\begin{tabular}[h!]
\renewcommand{\arraystretch}{1.5}
\centering
%\rotatebox{90}{
%\Rotatebox{90}{
%\begin{landscape}
%caption={Your table caption},
%label={tab:mytable},
%botcap, % caption below table
%sideways % This rotates the table
%]
%\begin{tabular*}{\textwidth}{c | c c  c c} 
%\begin{tabular*}{ l p{7cm} }{c | c c  c c} 
%\begin{tabular*}{width=\textwidth}{c | c c  c c} 
%\begin{tabular*}{1.2\textwidth}{c | c c  c c} 
\begin{tabular*}{1.3\textwidth}{c | c c  c c}
%\begin{tabular*}{1.5\textwidth}{c | c c  c c} 
\br
 Quantity & 
   Weyl scalar &
    Hertz potential &
  Corrector 
  &
 Metric \\ 
  \br
 Field eq. & 
 $\mathcal{O} \psi_0 = {}_{+2}T$ & 
$-\frac{1}{4}\thorn^4 \Phi = \bar{\psi}_0$&
 %$(\E x)\, l = T\, l$
$(\E x)_{ab}l^b = T_{ab}l^b$
 &
% $\E h = T$
$(\E h)_{ab} = T_{ab}$
 \\ \mr
  GF eq. & 
  $\O \, {}^\pm\mathcal{G}%(y,y_{\rm p}) 
  = \delta
  %(y,y_{\rm p})
  $
   & 
       $    -\tfrac{1}{4} \thorn^4 \, {}^\pm \mathcal{H}%(y,y_\textrm{p}) 
       =     {}^\pm\mathcal{G}%(y,y_\textrm{p})
       $
  &
 $l^b \, (\E {}^\pm {\mathcal X})_{ab}{}^{a_{\textrm{p}} b_{\textrm{p}}} \, l_{b_\textrm{p}} = g_a{}^{a_\textrm{p}} \delta$
% \left[ l^b \, (\E {}^\pm {\mathcal X})_{ab}{}^{a_{\textrm{p}} b_{\textrm{p}}} \, l_{b_\textrm{p}}\right] (y,y_\textrm{p}) = g_a{}^{a_\textrm{p}} \delta(y,y_{\textrm{p}})
  &
 % $\E^{\rm IRG}\, {}^\pm\mathcal{G}\stackrel{!}{=} g\, g\,\delta$
 $(\E^{\rm IRG}\, {}^\pm \mathcal{G})_{ab}{}^{a_\textrm{p} b_\textrm{p}}%(y,y_\textrm{p})
 \stackrel{!}{=} g_a{}^{a_\textrm{p}} g_b{}^{b_\textrm{p}}\delta
 %(y,y_\textrm{p})
 $
    \\ 
    (modes)
    &     &
    $-\tfrac{1}{4}  {\mathcal D}_{\omega m}^4
    {}^\pm h_{\omega \ell m}%(r,r_\textrm{p}) 
    = {}^\pm g_{\omega \ell m}$
    &&
    \\\mr
Field  
& 
$\psi^\pm_0
= {}^\pm\mathcal{G} *  {}_{+2}T$
& 
%$\Phi^\pm =  {}^\pm \bar {\mathcal H}* \bar \S\, T$
 $\Phi^\pm
    %(y) 
    =  {}^\pm \bar {\mathcal H}* \bar \S^{a_{\textrm{p}} b_{\textrm{p}}} T_{a_{\textrm{p}} b_{\textrm{p}}} 
    %(y)
    $
&
 %$x^\pm = {}^\pm {\mathcal X} * T$
       $ x_{ab}^\pm
       %(y) 
    = {}^\pm {\mathcal X}_{ab}{}^{a_{\textrm{p}} b_{\textrm{p}}} * T_{a_{\textrm{p}} b_{\textrm{p}}}
    %(y)
    $
&
%$h^\pm = \, {}^\pm \mathcal{G}* T$
$ h^\pm_{ab} = \, {}^\pm \mathcal{G}_{ab}{}^{a_\textrm{p} b_\textrm{p}}
    * T_{a_\textrm{p} b_\textrm{p}}$
\\

&&&&
%$h_{ab} = \dot g_{ab} + \Re(\S^\dagger \Phi)_{ab} + x_{ab} + ({\mathcal L}_X g)_{ab}$  
$h_{ab}$ in~\eqref{eq:decompi}
\\\mr
GF  
& 
${}^\pm\mathcal{G}$ in~\eqref{eq:Gretadv}
& 
${}^\pm\mathcal{H}$ in~\eqref{eq:Hdef}
&
${}^+ {\mathcal X}$ in~\eqref{eq:Xdef}
&
${}^\pm \mathcal{G}$ in ~\eqref{eq:GF MP}
\\ 
(modes)
&
${}^{\pm} g_{\omega \ell m}$ in \eqref{eq:g-}, \eqref{eq:g+}
&
%${}^\pm h$
${}^\pm h_{\omega \ell m}$
in  
\eqref{eq:h+1},\eqref{eq:h-1}
&& \\
\br
\end{tabular*}
%}
%\label{table:summary}
\end{adjustbox}
%\end{landscape}
%\rotatebox{90}{
%\caption{Summary for various field quantities of interest of the corresponding: field equation (Eqs.~\eqref{eq:Teuk psi0}, \eqref{eq:th4}, \eqref{eq:corr eq} and \eqref{eq:Eh=T}),  Green's function (GF) equation (Eqs.~\eqref{eq:GF eq psi0}, \eqref{eq:GP4H} and \eqref{eq:DGP}, \eqref{eq:GF eq corr} and Remark 4 within this subsection),  expression for the field (in terms of a convolution of the GF with the source; Eqs.~\eqref{psi0}, \eqref{propa:2}, \eqref{propa:1} and \eqref{eq:convolution}) and equation number containing an expression for the GF. For compactness and clarity, here we omit the
%%tensorial indices and the 
%spacetime points $y$ and $y_\textrm{p}$ in the  equations (we refer the reader to the equations in the main text for the expressions including the spacetime points).
%%full, detailed version of the equations).
%The symbol  $\stackrel{!}{=}$ means that the equality  only holds when applied to a conserved symmetric covariant rank-two tensor.}
%}
%\label{table:summary}
%\end{table}
%\end{tabular}
%}
%\end{landscape}
%\end{sidewaystable}
%\end{adjustbox}

\subsection{Propagation of singularities in IRG}
\label{sec:propsing}

Let us make a few final remarks comparing our advanced and retarded Green's functions in IRG ($h_{ab}l^b=0$) with those in the Lorenz gauge ($\nabla^a (h_{ab} - \frac{1}{2} hg_{ab})=0$)
regarding the propagation of singularities. This may be characterized in terms of wave front sets, as described in \cite{hormander1,hormander2} (to which we refer for mathematical details). If $S$ is a distribution on a manifold, we say that a point 
$(x,k_a)$ is in $WF(S)$, the wave front set of $S$, if $k_a \neq 0$ and if the Fourier transform of $\chi S$, where $\chi$ is any smooth function of compact support such that $\chi(x) \neq 0$, 
does not decay faster than any inverse power of $\lambda$ in the direction $\lambda k_a$ as $\lambda \to \infty$ in a local chart covering $x$, no matter how small the 
support of the window function $\chi$ is chosen. If there is no $(x,k_a)$ in the wave front set $WF(S)$ for a given 
point $x$, then $S$ is smooth in some neighborhood of $x$; otherwise, it is non-smooth at $x$ and the wave front set encodes the singular behavior in momentum space (infinitesimally) 
near $x$.

Let ${\mathcal P}$ be a scalar second-order partial differential operator on a globally hyperbolic spacetime $(\mathscr{M},g_{ab})$ whose highest derivative part is $\nabla^a \nabla_a$. Such an operator is strongly hyperbolic in the sense of \cite{Reula}, implying in particular that it has unique advanced and retarded 
Green's functions ${}^\pm {\mathcal K}(x,x')$, which are (bi-)distributions on $\mathscr{M} \times \mathscr{M}$. The propagation-of-singularities theorem for this operator $\mathcal P$ may be 
stated as 
\begin{equation}
\label{Psing}
\begin{split}
     WF({}^\pm {\mathcal K}) =& \ \{ (x,k_a,x',-k_a') \ : \ x \in J^\pm(x'), (x,k_a) \sim (x',k_a')\} \\
     \cup & \ \{(x,k_a,x,-k_a)\}.   
\end{split}
\end{equation}
Here, the notation $(x,k_a) \sim (x',k_a')$ means that $x$ and $x'$ are connected by a null geodesic curve $\xi$, such that $k^a$ is tangent to $\xi$ at $x$
and such that it is parallel transported to $k^{\prime a}$ at $x'$ along $\xi$. The wave front calculus tells us about the wave front set of the retarded or advanced solutions, 
$U^\pm = {}^\pm {\mathcal K} * S$, to the equation ${\mathcal P}U = S$ sourced by a distribution $S$ (of compact support, say):  
$WF({}^\pm {\mathcal K} * S)$ consists at most of such $(x,k_a)$ having the property that there is a $(x,k_a,x',-k_a')$ in $WF({}^\pm {\mathcal K})$ such that 
$(x',k_a') \in WF(S)$. Thus, the singularities -- in the sense of wave front set -- of $S$ get propagated in a specific way. For example, let 
\begin{equation}
    S(x) = \int \mu(\tau) \delta(\gamma(\tau),x) \, {\rm d}\tau,
\end{equation}
where $\gamma$ is some timelike curve and $\mu(\tau)$ is a smooth ``mass density'' of compact support. Then $WF(S)$ is contained in the set of 
$(\gamma(\tau),k_a)$ such that $\mu(\tau)\neq 0$ and such that $k_a \dot \gamma^a(\tau)=0$, i.e. the wave vectors $k_a$ are normal to $\gamma$, hence spacelike.
By \eqref{Psing} and the above rule of wave front calculus, the only elements of $WF({}^\pm {\mathcal K} * S)$ are in this case again
$(\gamma(\tau),k_a)$ such that $\mu(\tau)\neq 0$ and such that $k_a \dot \gamma^a(\tau)=0$. In particular, the wave front set is empty away from $\gamma$, so 
the solutions ${}^\pm {\mathcal K} * S$ are smooth away from $\gamma$. In other words, singularities are not propagated off of the worldline. These results apply, in particular, to the (adjoint) Teukolsky operator $\O \ (\O^\dagger)$. 

Let $\E^{\rm L}$ be the operator obtained from the linearized Einstein operator when we drop all terms that are zero in the Lorenz gauge. The crucial feature of the Lorenz gauge is that $\E^{\rm L} \propto \nabla^a \nabla_a +$ lower derivative terms. Even though $\E^{\rm L}$ is not a scalar operator, because the highest derivative part is a wave operator and ``diagonal'', the above discussion applies if we interpret the wave front set of a tensorial distribution as the union of the wave front sets of its components (in any chart). In particular, the retarded and advanced solutions to the EEs in Lorenz gauge sourced by the stress tensor of a particle on a timelike worldline have their singularities confined to the worldline. 

Next, 
let $\E^{\rm IRG}$ be the operator obtained from the linearized Einstein operator when we drop all terms that are zero in IRG (the explicit form of this operator can be obtained from the GHP form of $\E$ given, e.g., in \cite{Green:2019nam}). Different from the Lorenz gauge, $\E^{\rm IRG}$ is not strongly hyperbolic, so the standard solution theory for such systems \cite{Reula} does not apply. This does not preclude the existence of causal Green's functions in IRG, and indeed we have constructed them. Let us ask what is the propagation of singularities for these Green's functions
${}^\pm {\mathcal G}_{ab}{}^{a'b'}(x,x')$. 
Unlike in the Lorenz gauge, their wave front set does not straightforwardly follow from known general theorems, but we may determine it from the individual pieces on the right side of 
\eqref{eq:finalprop}.

Firstly, the distributions ${}^\pm {\mathcal X}_{ab}{}^{a'b'}(x,x')$ are the retarded respectively advanced
Green's functions of the GHZ transport equations along outgoing principal null geodesics tangent to $l^a$. It may be seen from this and the propagation-of-singularities theorem applied to transport operators that $(x,k_a, x', -k_a')$ is in $WF({}^\pm {\mathcal X})$ if and only if 
$x$ and $x'$ are on the same outgoing principal null geodesic tangent to $l^a$ (with $x'$ to the past/future of $x$), such that $l^ak_a=0$, and such that $k_a$ is parallel transported to $k_a'$ along that 
null geodesic. However, no condition that $k^a$ be null is required. We denote this relation by $(x,k_a) \approx (x',k_a')$, and with this notation
\begin{equation}
\label{Xsing}
\begin{split}
     WF({}^\pm {\mathcal X}_{ab}{}^{a'b'}) =& \ \{ (x,k_a,x',-k_a') \ : \ x \in J^\pm(x'), (x,k_a) \approx (x',k_a')\} \\
     \cup & \ \{(x,k_a,x,-k_a)\}.   
\end{split}    
\end{equation}

Let us next consider the term $ {\rm Re} \left[ {\mathcal S}^\dagger_{ab} \, {}^\pm \overline{\mathcal{H}}(x,x') \, 
       \overline{\mathcal{S}}^{a'b'} \right]$
on the right side of 
\eqref{eq:finalprop}. The wave front set of this term can be analyzed as follows. 
It is a general fact that differential operators such as ${\mathcal S}$ and ${\mathcal S}^\dagger$ cannot enlarge the wave front set. The kernels ${}^\pm \mathcal{H}(x,x')$
are related to the causal Green's functions ${}^\pm \mathcal{G}(x,x')$ of the Teukolsky operator by \eqref{eq:GP4H} and ${}^\pm \mathcal{G}(x,x')$ itself satisfies \eqref{Psing}.
If ${}^\pm {\mathcal I}(x,x')$
are the retarded and advanced propagators of the operator $\thorn^4$, we can write 
$-4{}^\pm {\mathcal I} * {}^\pm {\mathcal G} = {}^\pm {\mathcal H}$ by  \eqref{eq:GP4H}. The wave front calculus for composition $*$ alluded to above, and the fact that 
$WF({}^\pm {\mathcal I})$
is the same as in \eqref{Xsing} can be seen to imply the final result for the propagation-of-singularities in IRG:
\footnote{We strongly expect to have $=$ instead of inclusion $\subset$ below in practice.}
\begin{equation}
\label{Gsing}
\begin{split}
     WF({}^\pm {\mathcal G}_{ab}{}^{a'b'}) \subset& \ \{ (x,k_a,x',-k_a') \ : \ x \in J^\pm(x'), (x,k_a) \sim (x',k_a')\} \\
     \cup & \ \{(x,k_a,x,-k_a)\} \\   
     \cup & \ \{ (x,k_a,x',-k_a') \ : \ x \in J^\pm(x'), (x,k_a) \approx (x',k_a')\}.
\end{split}    
\end{equation}
The last  line is new compared to the corresponding result in Lorenz gauge \eqref{Psing}. 
It means that $(x',k_a')$, with $l^a(x')k_a'=0$ but $k_a'$ not necessarily null, can propagate along an outgoing principal null geodesic, 
into the causal future, respectively past of $x'$.

As a consequence of the last line, the retarded or advanced IRG solution $h^\pm_{ab}$ to the EEs with a source of the form 
\begin{equation}
\label{DetweilerSET}
    T^{ab}(x) = m \int  \dot \gamma^a \dot \gamma^b \delta(\gamma(\tau),x) \, {\rm d}\tau
\end{equation}
will have a wave front set of the form
\begin{equation}
    WF(h_{ab}^\pm) \subset 
    \{ (x,k_a) \ : \ 
    \exists (x',-k_a') \approx (x,k_a), x \in J^\pm(x'), x'=\gamma(\tau), \dot \gamma^a(\tau)k_a'=0\}.
\end{equation}
Thus, these solutions are  singular along future respectively past directed outgoing principal null 
geodesics intersecting the worldline $\gamma$. Such ``string singularities'' have previously been analysed in IRG by \cite{Pound:2013faa,HT2}, though 
their strength was examined by \cite{Pound:2013faa,HT2}  quantitatively and in position space, and not from the point of view of wave front techniques.

\section{Conclusions}

In this paper we have given advanced and retarded Green's functions for gravitational perturbations of Kerr in IRG
($h_{ab}l^b = 0$). Our method relied on the well-known metric reconstruction procedure, which has to be completed by a term in the Green's functions supported along outgoing principal null geodesics. The reconstructed part of the Green's functions is constructed in the frequency domain using (homogeneous) mode solutions of Teukolsky by a procedure due to Ori \cite{Ori:2002uv}. By contrast, the completion term -- the essential innovation of the present paper -- is constructed in the time domain using the GHZ procedure \cite{Green:2019nam}. It would be interesting to see if our procedure could be applied to Green's functions appearing in quantum field theory, notably the time-ordered (Feynman) Green's function. It would also be interesting for self-force calculations to see whether an analog of the Detweiler-Whiting Green's function \cite{DetweilerWhiting} could be obtained in IRG with our method. This would require a better understanding of the reconstructed part in the time domain.

\medskip
{\bf Acknowledgements:} S.H. thanks the Erwin Schrödinger Institut in Vienna, where part of this work has been completed, for its hospitality and support.
S.H. is grateful to the Max-Planck Society for supporting the collaboration between MPI-MiS and Leipzig U., grant Proj. Bez. M.FE.A.MATN0003. A.P. acknowledges the support of a Royal Society University Research Fellowship and the ERC Consolidator/UKRI Frontier Research Grant GWModels (as selected by the ERC and funded by UKRI [grant number EP/Y008251/1]). Many of the calculations in this work were enabled by the xAct \cite{xact1,xact2} tensor algebra package for Mathematica.

%%%%%%%%%%%%%%%%%%%
%%%%%%%%%%%%%%%%%%%

\appendix

\section{GHZ transport equations \cite{Green:2019nam}}

The corrector tensor field $x_{ab}$ is determined in such a way that
\begin{equation}\label{eq:x}
[T_{ab} - (\E x)_{ab}] l^b = 0, 
\end{equation}
with the idea to eliminate any $l$ component from $T_{ab}$. The ansatz for $x_{ab}$ was
\begin{equation}\label{eq:xdef1}
x_{ab} = 2m_{(a} \bar{m}_{b)} x_{m\bar{m}} -2l_{(a} \bar{m}_{b)} x_{nm} -2l_{(a} m_{b)} x_{n\bar{m}} + l_a l_b x_{nn},
\end{equation}
where $x_{n\bar{m}} = \bar{x}_{nm}$, so there are 4 real independent components encoded in $x_{m\bar{m}}$, $x_{mn}$ and $x_{nn}$ by which we attempt to satisfy the 4 real independent equations \eqref{eq:x}. We now first transvect \eqref{eq:x} with $l^a$ and use the $ll$ component of the Einstein operator $\E$ (see e.g. the appendix in~\cite{Green:2019nam} for the linearized EEs in GHP), to obtain
\begin{equation}
\label{eq:xmmb1}
\begin{split}
\left[ \thorn(\thorn-\rho-\bar{\rho}) +2\rho \bar{\rho} \right] x_{m\bar{m}} &= T_{ll}.
\end{split}
\end{equation}
Next, we transvect \eqref{eq:x} with $m^a$ and use the $ml$ component of the Einstein operator, to obtain
\begin{equation}\label{eq:xnm1}
\begin{split}
& \half\left[\thorn(\thorn-2\rho) + 2\bar{\rho}(\rho-\bar{\rho})\right]x_{nm}\\
=&\ T_{lm} - \half\bigg[(\thorn+\rho-\bar{\rho})(\eth+\bar{\tau}'-\tau) + 2\bar{\tau}'(\thorn-2\rho) -
(\eth-\tau-\bar{\tau}')\bar{\rho} +2\rho\tau\bigg]x_{m\bar{m}}.
\end{split}
\end{equation}
Finally, we transvect \eqref{eq:x} with $n^a$ and use the $nl$ component of the Einstein operator, to obtain
\begin{equation}
\label{eq:xnn1}
\begin{split}
 &\half\left[\rho(\thorn-\rho) + \bar{\rho}(\thorn-\bar{\rho})\right]x_{nn}\\
 =
 &\ T_{ln}- \half\bigg[(\eth'+\tau'-\bar{\tau})(\eth-\tau+\bar{\tau}') +
(\eth'\eth-\tau\tau'-\bar{\tau}\bar{\tau}'+\tau\bar{\tau}) - (\Psi_2+\bar{\Psi}_2)\\
&\pheq +(\thorn'-2\rho')\bar{\rho} + (\thorn-2\bar{\rho})\rho' +\rho(3\thorn'-2\bar{\rho}')
+\bar{\rho}'(3\thorn-2\rho)\\
&\pheq -2\thorn'\thorn + 2\rho\bar{\rho}' +2\eth'(\tau)-\tau\bar{\tau}\bigg]x_{m\bar{m}}
 \\
&\pheq - \half\bigg[(\thorn-2\rho)(\eth'-\bar{\tau}) + (\tau'+\bar{\tau})(\thorn+\bar{\rho})
-2(\eth'-\tau')\rho-2\bar{\tau}\thorn \bigg]x_{nm}\\
&\pheq - \half\bigg[(\thorn-2\bar{\rho})(\eth-\tau) + (\bar{\tau}'+\tau)(\thorn+\rho)
-2(\eth-\bar{\tau}')\bar{\rho}-2\tau\thorn \bigg]x_{n\bar{m}}.\\
\end{split}
\end{equation}
The operators on the left of these transport equations can be rewritten as in 
\eqref{eq:xmmb}, \eqref{eq:xnm} and \eqref{eq:xnn} above.

\section{Zero modes in IRG \cite{HT2}}\label{app:D}

A zero mode is any metric perturbation that is gauge equivalent to the perturbation
\begin{equation}\label{dgZ}
h^Z_{cb} = \delta M\, \partial_M g^{M,a}_{cb} + \delta a\, \partial_a g^{M,a}_{cb},
\end{equation}
with $g^{M,a}_{cd}$ the Kerr-family in some coordinate system (e.g., advanced KN or BL coordinates).
We define a particular representative of the zero mode by taking
\begin{equation}\label{dg}
\dot g_{ab} = h^Z_{ab} + {\mathcal L}_\zeta g_{ab},
\end{equation}
with a gauge vector $\zeta^a$ chosen in such a way that $\dot g_{ab}$ is in an IRG, namely $\dot g_{ab}l^b = 0 = \dot g_{ab} m^a \mb^b$. As shown 
in \cite{HT2}, we have
\begin{equation}
\zeta_b = \frac{2 \delta a}{a} \Re \left( \tau^\circ \bar{\tau}^\circ \rho\, l_b - \bar{\tau}^\circ m_b \right) ,
\end{equation}
using Held's notations reviewed in appendix
\ref{sec:Held}.
The components of $\dot g_{ab}$ are 
\begin{equation}
\begin{split}
\dot g_{nn} =& \left( \rho + \bar{\rho} \right) g^\circ_M + 2 \rho \bar{\rho} \left( \rho + \bar{\rho} \right) \Re \left(\half \Omega^\circ \tilde{\eth}' g^\circ_a + \bar{\tau}^\circ g^\circ_a \right),\\
\dot g_{nm} =& \rho \left( \rho + \bar{\rho} \right) g^\circ_a,
\label{zeromode1}
\end{split}
\end{equation}
and 
\begin{equation}
\dot g_{mm} = \dot g_{nl} = \dot g_{ml} = \dot g_{ll} = \dot g_{\mb m} = 0,
\label{zeromode2}
\end{equation}
where $g^\circ_M = - 2 \rho^{\prime \circ} \Psi^\circ \frac{\delta M}{M} \circeq \left\{ -3, -3 \right\}$ and $g^\circ_a = - \tau^\circ \Psi^\circ \frac{\delta a}{a} \circeq \left\{ -2, -4 \right\}$.

\section{Held's variant of GHP \cite{Held}}
\label{sec:Held}

While the GHP formalism is the most efficient for manipulating geometric equations in Kerr, it can be advantageous to work with a different basis of directional derivative operators in situations where one wants to specifically analyze the ``$r$''-dependence of quantities in retarded KN coordinates $(u,r,\theta,\varphi_*)$ (see equation \eqref{eq:KNdef}) in a manifestly GHP invariant fashion. $r$ itself is just a coordinate and certainly also not GHP invariant. But we have at our disposal the closely related, invariantly defined GHP scalar  $\rho$, 
\begin{equation}
\label{eq:Heldrho}
\frac{1}{\rho} = - r + ia \cos \theta
\quad \text{(Kinnersley frame)}.
\end{equation}
Different from $r$, it is complex -- and of course not a coordinate -- though it is equal to $-r$ in Schwarzschild (where $a=0$). Due to the type D relation 
\begin{equation}
\thorn \frac{1}{\rho} = -1,
\end{equation}
the operator $\thorn$ is formally like a partial derivative ``$\thorn = \partial_{-1/\rho}$'', and this formal relation is the essence of the Held formalism \cite{Held}. Of course $\thorn$ is not really a coordinate derivative as $1/\rho$ is not a genuine coordinate. In particular, we have no coordinate derivatives commuting with $\thorn$ either, closely related to the fact that, in Kerr, the distribution of 2-subspaces orthogonal to the tangential directions $n^a$ and $l^a$ is not integrable, i.e. is not tangent to a 2-dimensional surface. 
But it turns out that there are three preferred new operators which behave to a considerable extent like coordinate derivatives associated with the other directions, as we now explain in some more detail.

First we make a definition: We use the circle notation as in $x^\circ$ to mean that (i) $x^\circ$ is a GHP scalar of some definite weights and (ii) $\thorn x^\circ = 0$, i.e. 
informally, it does not depend on the ``coordinate'' $1/\rho$. The choice of $1/\rho$ as the ``coordinate'' as opposed to ``$1/\rho'$'' breaks the democracy between $l^a$ and $n^a$ manifest in the GHP formalism so the method of Held is appropriate when one of $l^a$ or $n^a$ is singled out, for example by the boundary conditions for the differential relations to be considered or by calculations in the IRG. 

Held's formalism \cite{Held} has three main ingredients. The first is a set of new operators, denoted by $\tilde{\eth}, \tilde{\eth}'$ and $\tilde{\thorn}'$ which replace $\eth, \eth'$ and $\thorn'$ and contrary to the latter have the property that $\tilde{\eth} \; x^\circ, \tilde{\eth}' x^\circ$ and $\tilde{\thorn}' x^\circ$ are quantities annihilated by $\thorn$.
The second is a list of identities, generated from the EEs and Bianchi identities, expressing any background quantity in terms of $\rho$ and quantities annihilated by $\thorn$, together with a table of how the new operators act on these. The third ingredient is an \emph{unambiguous} expansion of GHP scalars as Laurent polynomial in $\rho$ with coefficients annihilated by $\thorn$, i.e. we have the following:

\begin{lemma}
\label{lem:HeldFund}
Consider finitely many GHP quantities $a^\circ_{n}$ annihilated by $\thorn$ such that 
\begin{equation}\label{eq:poly1}
x=\sum^N_{n = 0} a^\circ_n \rho^{-n} = 0,
\end{equation}
for some $N \in \mathbb{N}$, in an open neighborhood of $\mathscr{M}$, where $x$ is a frame independent GHP scalar (i.e. it has a definite GHP weight). Then $a^\circ_n = 0 \quad \forall \quad 0 \leq n \leq N$.
\end{lemma}

We begin by recalling the precise definition of the new operators \cite{Held} on weight $\GHPw{p}{q}$ GHP quantities,
\begin{subequations}\label{eq:Hops}
\begin{align}
\tilde{\thorn}' &=\thorn' - \bar{\tau} \tilde{\eth} - \tau \tilde{\eth}' + \tau \bar{\tau} \left( \frac{q}{\rho} + \frac{p}{\bar{\rho}} \right) + \half \left( \frac{q \bar{\Psi}_2}{\bar{\rho}} + \frac{p \Psi_2}{\rho} \right), \\
\tilde{\eth} &=\frac{1}{\bar{\rho}} \eth + \frac{q\tau}{\rho}, \\
\tilde{\eth}' &= \frac{1}{\rho} \eth' + \frac{p\bar{\tau}}{\bar{\rho}}.
\end{align}
\end{subequations}
Using the vacuum EEs in GHP form, \cite{Held} next shows that on a type D background of the Case II in  Kinnersley's classification~\cite{Kinnersley:1969zza} (which includes Kerr) with a null tetrad aligned with the principal null directions,
\begin{subequations}\label{eq:exp}
  \begin{align}
    \rho' 
    %&= \rho^{\prime \circ} \bar{\rho} - \half \Psi^\circ \rho^2 - (\tilde{\eth} \bar{\tau}^\circ + \half \Psi^\circ) \rho \bar{\rho} - \tau^\circ \bar{\tau}^\circ \rho^2 \bar{\rho}\\
    &= \half (\rho + \bar{\rho}) \rho^{\prime \circ} + \half (\rho - \bar{\rho}) \bar{\rho}^{\prime \circ} - \half \rho^2 \Psi^\circ - \half \rho \bar{\rho} \bar{\Psi}^\circ - \rho^2 \bar{\rho} \tau^\circ \bar{\tau}^\circ,\\
    \tau' &= -\bar{\tau}^\circ \rho^2, \\
    \tau &= \tau^\circ \rho\bar\rho, \\
    \Psi_2 &= \Psi^\circ \rho^3,
  \end{align}
\end{subequations}
with $\tau^\circ \circeq \GHPw{-1}{-3}$, $\Psi^\circ \circeq \GHPw{-3}{-3}$, $\rho^{\prime \circ} \circeq \GHPw{-2}{-2}$, together with 
\begin{equation}
\label{eq:barrho}
\Omega^\circ = \frac{1}{\bar{\rho}} - \frac{1}{\rho} \circeq \GHPw{-1}{-1}.
\end{equation}
Consequently,
\begin{subequations}\label{eq:Hops2}
\begin{align}
\thorn' &= \tilde{\thorn}' + \rho \bar{\rho} \bar{\tau}^\circ \tilde{\eth} + \rho \bar{\rho} \tau^\circ \tilde{\eth}' - \rho \bar{\rho} \tau^\circ \bar{\tau}^\circ \left( p \rho + q\bar{\rho} \right) - \half \left( p \Psi^\circ \rho^2 + q \bar{\Psi}^\circ \bar{\rho}^2 \right), \\
\eth &= \bar{\rho} \tilde{\eth} - q \bar{\rho}^2 \tau^\circ, \\
\eth' &= \rho \tilde{\eth}' - p \rho^2 \bar{\tau}^\circ,
\end{align}
\end{subequations}
acting on GHP scalar with the weight of $\GHPw{p}{q}$.
\begin{table}[t]\label{DerivativeofHeldQuantities}
\begin{indented}
\item[]
\begin{tabular}{ c | c  c  c }
\br
& $\tilde{\thorn}'$ & $\tilde{\eth}$ & $\tilde{\eth}'$ \\ \mr
$\rho$ & 
$\rho^2 \rho^{\prime \circ} - \half \rho^2 (\rho \Psi^\circ + \bar{\rho} \bar{\Psi}_2^\circ) - \rho^3 \bar{\rho} \tau^\circ \bar{\tau}^\circ$ &
$\tau^\circ \rho^2$ &
$-\bar{\tau}^\circ \rho^2$ \\
$\Omega^\circ$ & $0$ & $2\tau^\circ$ & $-2\bar{\tau}^\circ$\\ 
$\rho^{\prime \circ}$ & $0$ & $0$ & $0$ \\
$\tau^\circ$ & $0$ & $0$ & 
                       $\half(\rho^{\prime \circ} + \bar{\rho}^{\prime \circ})\Omega^\circ + \half(\Psi^\circ - \bar{\Psi}^\circ)$\\
$\Psi^\circ$ & $0$ & $0$ & $0$ \\
\br
\end{tabular}
\end{indented}
\caption{Action of Held's operators on GHP background quantities in type D spacetimes of Kinnersley's Case II, in a tetrad aligned 
with the principal null directions \cite{Held}.} \label{tab:1}
\end{table}

The commutators between Held's operators read
\begin{equation}
\begin{split}
[\thorn, \tilde{\thorn}'] &= \left( \frac{\bar{\tau} \bar{\tau}'}{\bar{\rho}} + \frac{\tau\tau'}{\rho} - \half  \frac{\Psi_2}{\rho} - \half  \frac{\bar{\Psi}_2}{\bar{\rho}} \right) \thorn\\
&= \left(\rho \bar{\rho} (\rho + \bar{\rho}) \tau^\circ \bar{\tau}^\circ + \half \rho^2 \Psi^\circ + \half \bar{\rho}^2 \bar{\Psi}^\circ \right) \thorn,  \\
[\thorn, \tilde{\eth}] &= -\left( \frac{\tau}{\rho} + \frac{\bar{\tau}'}{\bar{\rho}} \right) \thorn 
= 0, 
\end{split}
\end{equation}
which express that both $\tilde{\eth}, \tilde{\thorn}'$ when applied to a quantity $f^\circ$ annihilated by $\thorn$ give another such quantity, as well as
\begin{equation}
\begin{split}
[\tilde{\eth}, \tilde{\eth}'] &= \frac{\bar{\rho}'-\rho'}{\rho\bar{\rho}} \thorn + \Omega^\circ \tilde{\thorn}' + p\Sigma^\circ - q \bar{\Sigma}^\circ{}'\\
&= \left( - \frac{\rho^{\prime \circ}}{\bar{\rho}} + \frac{\bar{\rho}^{\prime \circ}}{\rho} + \half \rho \Omega^\circ \Psi^\circ + \half \bar{\rho} \Omega^\circ \bar{\Psi}^\circ + \rho \bar{\rho} \Omega^\circ \tau^\circ \bar{\tau}^\circ \right) \thorn + \Omega^\circ \tilde{\thorn}' + p \rho^{\prime \circ} - q \bar{\rho}^{\prime \circ},\\
[\tilde{\thorn}', \tilde{\eth}'] &=- \frac{\bar{\tau}(\bar{\rho}'-\rho')}{\rho\bar{\rho}} \thorn + p\Gamma^\circ\\
&= \left( \rho \rho^{\prime \circ} \bar{\tau}^\circ - \bar{\rho} \bar{\rho}^{\prime \circ} \bar{\tau}^\circ - \half \rho^2 \bar{\rho} \bar{\tau}^\circ \Omega^\circ \Psi^\circ - \half \rho \bar{\rho}^2 \bar{\tau}^\circ \Omega^\circ \bar{\Psi}^\circ - \rho^2 \bar{\rho}^2 \Omega^\circ \tau^\circ \bar{\tau}^{\circ 2} \right) \thorn,
\end{split}
\end{equation}
where
\begin{equation}
\begin{split}
\Sigma^\circ &= \frac{\rho'}{\bar{\rho}} + \frac{\Psi_2}{2\rho} \left( \frac{1}{\rho} + \frac{1}{\bar{\rho}} \right) + \tilde{\eth} \left( \frac{\bar{\tau}}{\bar{\rho}} \right) = \rho^{\prime \circ},\\
\Gamma^\circ &= \tilde{\thorn}' \left( \frac{\bar{\tau}}{\bar{\rho}} \right) - \frac{\bar{\tau} \Psi_2}{\rho \bar{\rho}} - \frac{\bar{\tau} \rho'}{\bar{\rho}} -
\half \tilde{\eth}' \left( \frac{\Psi_2}{\rho} \right) = 0.
\end{split}
\end{equation}
These identities are complemented by identities for the action of the new operators on $\rho$ and on  $\Omega^\circ, \rho^{\prime \circ}, \tau^\circ, \Psi^\circ$ given in table \ref{tab:1}, which is extracted from \cite{Held} (and specialized to type D metrics such as Kerr).

In the Kinnersley tetrad and retarded KN coordinates, the Held scalars reduce to the entries of table \ref{tab:2}.
\begin{table}[t]\label{eq:Held coeffs in Kinn}%\addtolength{\tabcolsep}{3pt}
\renewcommand{\arraystretch}{1.5}
\begin{indented}
\item[]
\begin{tabular}{  c  c }
\br
Held quantity & Value in Kinnersley frame \\ \mr 
$\tau^\circ$ & $-\frac{1}{\sqrt{2}} ia\sin\theta$ \\
$\rho^{\prime\circ}$ & $-\frac{1}{2}$ \\
$\Psi^\circ$ & $M$ \\[-.5em]
$\Omega^\circ$ & $-2ia\cos\theta$ \\[-.5em]
$\thorn$ & $\partial_r$ \\
$\tilde{\thorn}'$ & $\partial_u - \frac{1}{2} \frac{r^2 - 2 M r + a^2}{r^2 + a^2 \cos^2 \theta} \partial_r$\\
$\tilde{\eth}$ & $-\frac{1}{\sqrt{2}} \left( \partial_\theta 
+ i \csc \theta \partial_{\varphi_*} -i a \sin \theta \partial_u- s \cot \theta \right)$\\
$\tilde{\eth}'$ & $-\frac{1}{\sqrt{2}} \left( \partial_{\theta} - i \csc \theta \partial_{\varphi_*} + ia\sin \theta \partial_u + s \cot \theta \right)$\\
\br
\end{tabular}
\end{indented}
\caption{Held's GHP scalars/operators in Kerr in retarded KN coordinates and the Kinnersley frame \cite{HT2}. 
One can recognize Held's operators $\tilde{\eth}$ and $\tilde{\eth}'$ as being closely related to standard operators in the theory of spin-weighted harmonics
appearing, e.g., in \cite{Chandrasekhar:1984siy}.} \label{tab:2}
\end{table}
The operators and relations in the Held formalism appear to be quite complicated looking but their use in fact strongly simplifies and 
conceptualizes any calculations involving GHP operators and expressions expanded in terms of $\rho$ and quantities 
that are annihilated by $\thorn$, as frequently appear in our analysis.

\section{Addition theorems for spin-weighted spherical harmonics}
\label{app:addthm}

With the normalization for the spin-weighted spherical harmonics given in \eqref{eq:SpheroidalHarmonicsIdentities}, one has the
following addition theorem:
\begin{equation}\label{eq:SpinWeightedClebschGordon}
\begin{split}
\Ylm{s_1}{j_1}{m_1} \Ylm{s_2}{j_2}{m_2} =& \sum_{s_3, j_3, m_3} (-1)^{s_3 + m_3} \sqrt{\frac{(2j_1+1)(2j_2+1)(2j_3+1)}{4\pi}} \times \phantom{
\begin{matrix}
j\\
m
\end{matrix}
} \\
& \left(
\begin{matrix}
j_1 & j_2 & j_3\\
m_1 & m_2 & -m_3
\end{matrix}
\right) \left(
\begin{matrix}
j_1 & j_2 & j_3\\
-s_1 & -s_2 & s_3
\end{matrix}
\right)
\Ylm{s_3}{j_3}{m_3},
\end{split}
\end{equation}
where
\begin{equation}
\left(
\begin{matrix}
j_1 & j_2 & j_3\\
m_1 & m_2 & m_3
\end{matrix}
\right)
\end{equation}
are the Wigner 3-$j$ symbols. One can prove this 
theorem by applying products of Chandrasekhar's raising and lowering operators \eqref{eq:Chandop} to the Clebsch-Gordan decomposition for the ordinary, un-weighted, spherical harmonics. 

The actions of $\tilde{\eth}$ and $\tilde{\eth}'$ (see table \ref{tab:2}) on a harmonic $e^{-i\omega u} {}_s Y_{\ell m}(\theta, \varphi_*)$ are best computed using 
the relation with the spin-raising and -lowering operators \eqref{eq:Chandop}, combined with the formulas
\begin{equation}\label{eq:RaisingSpinWeightedOpWithOmegaOnHarmonics}
\begin{split}
\left( \tilde{\eth}' - {}_s\mathcal{L}^\dagger_m \right) \Y{s} e^{-i\omega u} &= - a \omega \sqrt{\frac{4\pi}{3}} \; \Ylm{1}{1}{0} \; \Y{s} e^{-i\omega u} \\
&= a \omega \sum^{\ell+1}_{j=\ell-1} (-1)^{\ess+\emm} \sqrt{(2\ell+1)(2j+1)} \left(
\begin{matrix}
\ell & 1 & j\\
\emm & 0 & -\emm
\end{matrix}
\right) \times\\
&\quad \left(
\begin{matrix}
\ell & 1 & j\\
-s & -1 & s+1
\end{matrix}
\right) \Ylm{s+1}{j}{\emm} e^{-i\omega u} \\
&= a \omega \bigl( \preindex{\alpha}{s}^-_{\ell, \emm} \; \Ylm{s+1}{(\ell-1)}{\emm} + \preindex{\alpha}{s}^0_{\ell, \emm} \; \Ylm{s+1}{\ell}{\emm} \\
&\qquad\quad+ \preindex{\alpha}{s}^+_{\ell, \emm} \; \Ylm{s+1}{(\ell+1)}{\emm} \bigr) e^{-i\omega u},
\end{split}
\end{equation}
where
\begin{equation}
\begin{split}
\preindex{\alpha}{s}^-_{\ell, \emm} &= \frac{1}{\ell} \sqrt{\frac{(\ell-\emm)(\ell+\emm)(\ell-s)(\ell-s-1)}{2(2\ell+1)(2\ell-1)}} ,\\
\preindex{\alpha}{s}^0_{\ell, \emm} &= - \frac{\emm}{\ell (\ell+1)} \sqrt{\frac{(\ell-s)(\ell+s+1)}{2}},\\
\preindex{\alpha}{s}^+_{\ell, \emm} &= - \frac{1}{\ell+1} \sqrt{\frac{(\ell-\emm+1)(\ell+\emm+1)(\ell+s+1)(\ell+s+2)}{2(2\ell+1)(2\ell+3)}},
\end{split}
\end{equation}
as well as
\begin{equation}\label{eq:LoweringSpinWeightedOpWithOmegaOnHarmonics}
\begin{split}
\left( \tilde{\eth} - {}_s \mathcal{L}_m \right) \Y{s} e^{-i\omega u} &= - a \omega \sqrt{\frac{4\pi}{3}} \; \Ylm{-1}{1}{0} \; \Y{s} e^{-i\omega u}\\
&= a \omega \sum^{\ell+1}_{j=\ell-1} (-1)^{\ess+\emm} \sqrt{(2\ell+1)(2j+1)} \left(
\begin{matrix}
\ell & 1 & j\\
\emm & 0 & -\emm
\end{matrix}
\right) \times\\
&\quad \left(
\begin{matrix}
\ell & 1 & j\\
-s & 1 & s-1
\end{matrix}
\right) \Ylm{s-1}{j}{\emm} e^{-i\omega u} \\
&= a \omega \bigl( \preindex{\alpha}{-s}^-_{\ell, \emm} \; \Ylm{s-1}{(\ell-1)}{\emm} - \preindex{\alpha}{-s}^0_{\ell, \emm} \; \Ylm{s-1}{\ell}{\emm} \\
&\qquad\quad+ \preindex{\alpha}{-s}^+_{\ell, \emm} \; \Ylm{s-1}{(\ell+1)}{\emm} \bigr) e^{-i\omega u}.
\end{split}
\end{equation}
When evaluating the product of spin-weighted harmonics with the Held scalars $\Omega^\circ$ and $\tau^\circ$ (see table \ref{tab:2}), it is convenient to use the formulas
\begin{equation}
\begin{split}
\tau^\circ \Y{s} &= - i a \sqrt{\frac{4\pi}{3}} \; \Ylm{1}{1}{0} \; \Y{s}\\
&= i a \left( \preindex{\alpha}{s}^-_{\ell, \emm} \; \Ylm{s+1}{(\ell-1)}{\emm} + \preindex{\alpha}{s}^0_{\ell, \emm} \; \Ylm{s+1}{\ell}{\emm} + \preindex{\alpha}{s}^+_{\ell, \emm} \; \Ylm{s+1}{(\ell+1)}{\emm} \right),
\end{split}
\end{equation}
\begin{equation}
\begin{split}
\bar{\tau}^\circ \Y{s} &= - i a \sqrt{\frac{4\pi}{3}} \; \Ylm{-1}{1}{0} \; \Y{s}\\
&= i a \left( \preindex{\alpha}{-s}^-_{\ell, \emm} \; \Ylm{s-1}{(\ell-1)}{\emm} - \preindex{\alpha}{-s}^0_{\ell, \emm} \; \Ylm{s-1}{\ell}{\emm} + \preindex{\alpha}{-s}^+_{\ell, \emm} \; \Ylm{s-1}{(\ell+1)}{\emm} \right),
\end{split}
\end{equation}
\begin{equation}
\begin{split}
\Omega^\circ \Y{s} &= - 2 i a \sqrt{\frac{4\pi}{3}} \; \Ylm{0}{1}{0} \; \Y{s}\\
&= 2i a \left( \preindex{\beta}{s}^-_{\ell, \emm} \; \Ylm{s}{(\ell-1)}{\emm} + \preindex{\beta}{s}^0_{\ell, \emm} \; \Ylm{s}{\ell}{\emm} + \preindex{\beta}{s}^+_{\ell, \emm} \; \Ylm{s}{(\ell+1)}{\emm} \right),
\end{split}
\end{equation}
where
\begin{equation}
\begin{split}
\preindex{\beta}{s}^-_{\ell, \emm} &= \frac{1}{\ell} \sqrt{\frac{(\ell-\emm)(\ell+\emm)(\ell-s)(\ell+s)}{(2\ell+1)(2\ell-1)}} ,\\
\preindex{\beta}{s}^0_{\ell, \emm} &= - \frac{\emm s}{\ell (\ell+1)},\\
\preindex{\beta}{s}^+_{\ell, \emm} &= - \frac{1}{\ell+1} \sqrt{\frac{(\ell-\emm+1)(\ell+\emm+1)(\ell-s+1)(\ell+s+1)}{(2\ell+1)(2\ell+3)}}.
\end{split}
\end{equation}

\section{Green's function  for GHZ transport equation in Schwarzschild}
\label{app:Schw}

In this appendix we give the expression for the Green's function components in Eqs.~\eqref{propa:3}-- \eqref{propa:5} in the simpler case of Schwarzschild spacetime (for which $a=0$, and so, $\rho=-1/r$ and $\tau^\circ = \Omega^\circ = 0$). We also substitute the 
values $\Psi^\circ = M, \rho^{\prime \circ} = -1/2$, and the expression $\tilde{\thorn}' = \partial_u$, which are valid in the Kinnersley frame (even in Kerr); see table \ref{tab:2}. We obtain:

% remove brackets around \Theta's; make brackets of factor in middle squared  instead of curved; subindices 'p'; superindex '+' in \Theta make it unspaced out
% \text{rp} -> r_\textrm{p}

\begin{equation}
\label{propa:3Schw}
{}^+ X^{ab}_l=
-l^a l^b \Theta _\textrm{p}^+\left(\frac{r_\textrm{p}^2}{r}\right) \delta _\textrm{p}^{\circ },
%2 l^a l^b \Theta^+_\textrm{p} \left[ (\rho + \bar{\rho}) \frac{1}{(\rho_\textrm{p} + \bar{\rho}_\textrm{p})^2} \right] \delta^\circ_\textrm{p},
\end{equation}

%\begin{subequations}
\begin{align}
\label{propa:4Schw}
&{}^+ X^{ab}_n = 
 m^{(a} \bar{m}^{b)} \Theta^+_\textrm{p} \left[ \
\frac{r_\textrm{p} (r-r_\textrm{p})}{r} \
\right]\delta^\circ_\textrm{p} 
- \frac{1}{6} l^{(a} \bar{m}^{b)} \Theta^+_\textrm{p} \
\left[\frac{ (r-r_\textrm{p})^2 (r+2 r_\textrm{p})}{r^2} \right] \
\tilde{\eth} \delta^\circ_\textrm{p}+
\nonumber \\&
\frac{1}{8} l^a l^b \Theta^+_\textrm{p} \bigg[ r_\textrm{p} (5r+4 r_\textrm{p}) \bigg] \partial_u \delta^\circ_\textrm{p}
- \frac{1}{6} l^a l^b \Theta^+_\textrm{p} \bigg[\frac{(r-r_\textrm{p})^3}{r^2} \bigg] \tilde{\eth}' \tilde{\eth} \
\delta^\circ_\textrm{p}-
\nonumber\\&
\frac{1}{2} l^a l^b \Theta^+_\textrm{p} 
\bigg[\frac{ M r_\textrm{p} (r-r_\textrm{p})}{r^2} \
\bigg] \delta^\circ_\textrm{p}
+\frac{1}{2}  l^a l^b \Theta^+_\textrm{p}  r_\textrm{p}  \delta^\circ_\textrm{p},
\end{align}
%\end{subequations}

\begin{equation}
\label{propa:5Schw}
{}^+ X^{ab}_{\bar{m}} =
\frac{2}{3} l^{(a}\bar{m}^{b)}\Theta^+_\textrm{p}\left(
r-\frac{r_\textrm{p}^3}{r^2} \right)\delta^\circ_\textrm{p} 
+ \frac{1}{6} l^a l^b\Theta^+_\textrm{p}\left(\frac{2 r^3-3 r r_\textrm{p}^2+r_\textrm{p}^3}{r^2} \
\right)\tilde{\eth}\delta^\circ_\textrm{p}.
\end{equation}
In the above formulas, we could further substitute the expressions
\begin{equation}
\begin{split}
\tilde{\eth} &= -\frac{1}{\sqrt{2}} \left( \partial_\theta 
+ i \csc \theta \partial_{\varphi_*} - s \cot \theta \right),\\
\tilde{\eth}' &= -\frac{1}{\sqrt{2}} \left( \partial_{\theta} - i \csc \theta \partial_{\varphi_*}  + s \cot \theta \right),
\end{split}
\end{equation}
which are valid in Schwarzschild and the Kinnersley frame, noting that $\delta^\circ_{\textrm{p}}$ [see equation \eqref{eq:deltacirc}] has $s=0$.

%\marc{and we note that, even in Kerr, it is $\Psi ^{\circ }=M$ and $\rho^{\prime \circ}=-1/2$.}

%------------------------------------------------------------------------

\end{document}